\documentclass[sigconf, nonacm]{acmart}
\newcommand\vldbdoi{XX.XX/XXX.XX}
\newcommand\vldbpages{XXX-XXX}
\newcommand\vldbvolume{18}
\newcommand\vldbissue{5}
\newcommand\vldbyear{2025}
\newcommand\vldbauthors{\authors}
\newcommand\vldbtitle{\shorttitle} 
\newcommand\vldbavailabilityurl{\url{https://doi.org/10.5281/zenodo.14991194}}
\newcommand\vldbpagestyle{empty} 

\newif\iffull
\fulltrue
\newif\ifcomments
\commentstrue
\newif\ifmarknew
\marknewtrue
\newif\ifmarkrevised
\markrevisedfalse

\usepackage{microtype}
\usepackage{tikzscale}
\usepackage{xcolor}
\usepackage{graphicx}
\usepackage{pifont}
\usepackage{anyfontsize}

\usepackage{caption}
\usepackage{subcaption}
\usepackage{framed}
\usepackage{quoting}
\usepackage{xspace}
\usepackage[shortlabels]{enumitem}
\usepackage{cleveref}
\usepackage{proof}
\usepackage{balance}

\usepackage{listings}
\usepackage{algorithm}
\usepackage{algpseudocode}
\usepackage{stmaryrd}
\usepackage[framemethod=tikz]{mdframed}
\definecolor{backcolour}{rgb}{0.95,0.95,0.94}

\lstdefinelanguage{Isabelle}{
    keywords=[1]{definition, record, lemma, where, datatype, type_synonym, if, then, else, while, and},
    keywords=[2]{function, for, each, return, try:, except:, else:, in},
    keywords=[3]{acq_wr_lock, cl_prepare, cl_commit, cl_write_commit, cl_read_invoke, initiateTransaction, xview,prepare, commit, abort, acquireLocks, releaseLocks, tapir_occ_check, committed_wr_tstmps, prepared_wr_tstmps, prepared_rd_tstmps, get_ver_ts},
    alsoletter=':,
    sensitive=true, % keywords are not case-sensitive
    morecomment=[l]{--} % l is for line comment
}

\lstdefinestyle{Isabellestyle}{
    backgroundcolor=\color{backcolour}, 
    commentstyle=\color{olive}\rmfamily\itshape,
    keywordstyle=[1]\color{blue}\bfseries,
    keywordstyle=[2]\color{violet}\bfseries,
    keywordstyle=[3]\bfseries,
    numberstyle=\scriptsize\color{gray},
    basicstyle=\linespread{0.95}\ttfamily\small,
    basewidth=0.5em,
    breakatwhitespace=false,         
    breaklines=true,                 
    captionpos=b,
    columns=fixed,
    fontadjust=true,
    frame=single,
    keepspaces=true,                 
    mathescape,
    numbers=left,
    numbersep=5pt,
    rulecolor=\color{backcolour},                  
    showspaces=false,                
    showstringspaces=false,
    showtabs=false,                  
    tabsize=2,
}
\newtheorem{theorem}{Theorem}          % for global numbering

\ifcomments
\newcommand{\authorcomment}[3]{\textcolor{#1}{\textbf{[#2}: #3\textbf{]}}}
\else
\newcommand{\authorcomment}[3]{}
\fi

\ifmarknew

\else

\fi

\ifmarkrevised
\newcommand{\rev}[1]{\textcolor{darkred}{#1}}
\else
\newcommand{\rev}[1]{#1}
\fi

\newcommand{\inlsec}[1]{\smallskip\noindent\textbf{#1.}}

\newcommand{\ourframework}{\textsc{VerIso}\xspace}
\newcommand{\invcmts}{inverted commits\xspace}

\newcommand{\nats}{\mathbb{N}}

\newcommand{\fun}{\rightarrow}
\newcommand{\map}{\rightharpoonup}
\newcommand{\powset}[1]{\mathcal{P}(#1)}

\usepackage{xargs}  

\newcommandx{\myrightarrow}[1]{
  \xrightarrow{ \raisebox{-1.5pt}[2pt][0pt]{ \ensuremath{ \scriptstyle{#1} }}}
}

\newcommand{\Skip}{\mathsf{skip}}
\newcommand{\reach}{\mathsf{reach}}
\newcommand{\refines}{\ensuremath{\mathrel{\preccurlyeq}}}

\newcommand{\refmap}[2]{#1_{#2}}
\newcommand{\refmapfull}[3]{\refmap{#1}{#3}, \refmap{#2}{#3}}
\newcommand{\refmapkvs}[1]{\kvs\_\texttt{of}{#1}}
\newcommand{\refmapviews}[1]{\views\_\texttt{of}{#1}}

\DeclareTextFontCommand{\inlisa}{\ttfamily}   % for inline identifiers (type names, lemma names, ...)
\newcommand{\isaco}[1]{\inlisa{#1}}

\DeclareTextFontCommand{\isakw}{\rmfamily\bfseries}  % for Isabelle keywords: lemma, fun, definition, ....
\DeclareTextFontCommand{\isaid}{\rmfamily\itshape}   % for other identifiers (type names, lemma names, ...)

\newcommand{\isa}{\isaid}

\newcommand{\keytype}{\inlisa{key}}
\newcommand{\valuetype}{\inlisa{value}}
\newcommand{\verstype}{\inlisa{version}}

\newcommand{\listtype}[1]{\inlisa{list}(#1)}
\newcommand{\rwtype}{\{\inlisa{R,W}\}}

\newcommand{\txid}[2]{\ensuremath{\isaco{Tn}(#1,#2)}}

\newcommand{\SO}{\isaco{SO}}
\newcommand{\WW}{\isaco{WW}}
\newcommand{\WR}{\isaco{WR}}

\newcommand{\kvs}{\mathcal{K}}
\newcommand{\views}{\mathcal{U}}
\newcommand{\fprint}{\mathcal{F}}
\newcommand{\rfprint}{\fprint_{R}}
\newcommand{\wfprint}{\fprint_{W}}

\newcommand{\readonly}{\isaco{rdonly}}
\newcommand{\visTx}{\isaco{visTx}}

\newcommand{\LWW}{\isaco{LWW}}
\newcommand{\wellformed}{\isaco{wf}}

\newcommand{\canCommit}{\isaco{canCommit}}
\newcommand{\vShift}{\isaco{vShift}}

\newcommand{\isomodel}[1]{\mathcal{I}\!(#1)}

\newcommand{\tpl}{\inlisa{TPL}}

\newcommand{\isolevel}{\isa{IL}}
\newcommand{\RA}{\isaco{RA}}
\newcommand{\UA}{\isaco{UA}}
\newcommand{\TCC}{\isaco{TCC}}

\newcommand{\CP}{\isaco{CP}}
\newcommand{\PSI}{\isaco{PSI}}
\newcommand{\SI}{\isaco{SI}}
\newcommand{\SER}{\isaco{SER}}
\newcommand{\SSER}{\isaco{SSER}}

\begin{document}
\title{\ourframework: Verifiable Isolation Guarantees for Database Transactions}

%%
%% The "author" command and its associated commands are used to define the authors and their affiliations.
\author{Shabnam Ghasemirad}
\affiliation{%
  \institution{ETH Zurich, Switzerland}
  \country{}
  \city{}
}
\email{shabnam.ghasemirad@inf.ethz.ch}

\author{Si Liu}
\affiliation{%
  \institution{ETH Zurich, Switzerland}
  \country{}
  \city{}
}
\email{si.liu@inf.ethz.ch}

\author{Christoph Sprenger}
\affiliation{%
  \institution{ETH Zurich, Switzerland}
    \country{}
  \city{}
}
\email{sprenger@inf.ethz.ch}

\author{Luca Multazzu}
\affiliation{%
  \institution{ETH Zurich, Switzerland}
    \country{}
  \city{}
}
\email{lcmultazzu@gmail.com}

\author{David Basin}
\affiliation{%
  \institution{ETH Zurich, Switzerland}
    \country{}
  \city{}
}
\email{basin@inf.ethz.ch}

\begin{abstract}
\looseness=-1
Isolation bugs, stemming especially from design-level defects, have been repeatedly found in carefully designed and extensively tested production databases over decades. In parallel, various frameworks for modeling database transactions and reasoning about their isolation guarantees have been developed. What is missing however is a mathematically \emph{rigorous} and \emph{systematic} framework with tool support for formally verifying a wide range of such guarantees for \emph{all} possible system behaviors. We present the first such framework, \ourframework, developed within the theorem prover Isabelle/HOL. To showcase its use in verification, we model the strict two-phase locking concurrency control protocol and verify that it provides strict serializability isolation guarantee. Moreover, we show how \ourframework helps identify isolation bugs during protocol design. We derive new counterexamples for the TAPIR protocol from failed attempts to prove its claimed strict serializability. In particular, we show that it violates a much weaker isolation level, namely, atomic visibility.
\end{abstract}

\maketitle

\pagestyle{\vldbpagestyle}
\begingroup\small\noindent\raggedright\textbf{PVLDB Reference Format:}\\
\vldbauthors. \vldbtitle. PVLDB, \vldbvolume(\vldbissue): \vldbpages, \vldbyear.\\
\href{https://doi.org/\vldbdoi}{doi:\vldbdoi}
\endgroup
\begingroup
\renewcommand\thefootnote{}\footnote{\noindent
This work is licensed under the Creative Commons BY-NC-ND 4.0 International License. Visit \url{https://creativecommons.org/licenses/by-nc-nd/4.0/} to view a copy of this license. For any use beyond those covered by this license, obtain permission by emailing \href{mailto:info@vldb.org}{info@vldb.org}. Copyright is held by the owner/author(s). Publication rights licensed to the VLDB Endowment. \\
\raggedright Proceedings of the VLDB Endowment, Vol. \vldbvolume, No. \vldbissue\ %
ISSN 2150-8097. \\
\href{https://doi.org/\vldbdoi}{doi:\vldbdoi} \\
}\addtocounter{footnote}{-1}\endgroup

\ifdefempty{\vldbavailabilityurl}{}{
\vspace{.3cm}
\begingroup\small\noindent\raggedright\textbf{PVLDB Artifact Availability:}\\
The source code, data, and/or other artifacts have been made available at \vldbavailabilityurl
\endgroup
}

\section{Introduction}

Over the past decades, significant efforts have been devoted to developing reliable, high-performance databases. Starting from the centralized relational databases of the 1980s,  development progressed to geo-replicated NoSQL key-value stores such as Dynamo and Cassandra, prioritizing availability over consistency. This progress continued with the emergence of NewSQL databases, which bridge the gap between SQL and NoSQL by supporting ACID transactions for better data integrity, while still being highly performant.

On the theoretical side, this trend has led to the study of new isolation levels, accompanied by novel concurrency control mechanisms and database designs balancing isolation guarantees and system performance. In addition to the gold standard Serializability~(SER) as well as the widely adopted Snapshot Isolation (SI)~\cite{CockroachDB,MongoDB,Dgraph}, many weaker isolation levels have emerged to cater for real-world applications. These include Read Atomicity~(RA), supported by \cite{ramp,lora,noc-noc} and recently layered atop Facebook's TAO~\cite{RAMP-TAO}; Transactional Causal Consistency~(TCC), supported by~\cite{OCC:TPDS2021,FriendFoe:VLDB2018,Cure:ICDCS2016,Eiger:NSDI2013,SNOW:OSDI2016,noc-noc,UniStore:ATC2021,Slowdown:NSDI2017}, and recently adopted by commercial databases~\cite{CosmosDB,Neo4j,ElectricSQL}. 

%%
%% BUGS AND THE NEED FOR FORMAL METHODS
%%
\looseness=-1 Unfortunately, isolation bugs have repeatedly manifested themselves in carefully designed and heavily tested databases
\cite{jepsen-analyses,polysi,txcheck,DBLP:journals/pacmpl/BiswasE19,Troc:ICSE2023}.
In particular, all these 
%databases have exhibited 
isolation bugs stemmed from design-level defects, rather than 
%pure 
implementation errors. For example, Yugabyte\-DB was found to violate SER as its conflict-detection mechanism failed to account 
for read conflicts~\cite{YugaByte}, 
% MongoDB, resp. 
MariaDB's Galera cluster incorrectly claimed %its support for %full ACID transactions, resp. 
SI support~\cite{galera-issue-1,polysi}, and Lu et al.~\cite{osdi23} recently reported violations of the claimed Strict~SER~(SSER) in
TAPIR~\cite{DBLP:journals/tocs/ZhangSSKP18} and DrTM~\cite{DrTM}.
Eliminating such bugs, which requires costly redesigns, and achieving strong correctness guarantees is a crucial, challenging task for developers. Here, the formal verification of database designs is highly desirable, preferably in an early design stage. Indeed, the industrial adoption of formal methods in the wider area of production databases~\cite{amazon,mongoDB-TLA+,Azure,Apalache} has become increasingly prevalent as a proactive approach to address such design errors.

%%
%% FORMAL SEMANTICS
%%
The formal verification of concurrency control protocols necessitates a formal semantics for isolation guarantees. Many semantics have been proposed~\cite{bernstein1981concurrency, adya1999weak, DBLP:conf/concur/Cerone0G15, DBLP:journals/pacmpl/BiswasE19, seeing, DBLP:conf/tacas/LiuOZWM19, DBLP:conf/ecoop/XiongCRG19},
including the first characterization of SER via dependency graphs~\cite{bernstein1981concurrency}, an axiomatic framework based on abstract executions~\cite{DBLP:conf/concur/Cerone0G15}, and the recent operational semantics for isolation guarantees~\cite{DBLP:conf/ecoop/XiongCRG19}.
Nonetheless, protocol verification based on these semantics relies on either 
(i) manual pen-and-paper proofs, which are error-prone for complex protocols, 
(ii) on testing, which can only check a small number of executions, or (iii) on model checking, which is limited to a small number of processes and transactions, and can therefore find anomalies, but not establish correctness. Hence, these approaches can miss design errors that violate the claimed isolation guarantees. For example, 
%as mentioned above, 
TAPIR violates SSER, despite its authors' formal modeling  as well as both manual proofs and model checking efforts~\cite{DBLP:journals/tocs/ZhangSSKP18}. 
%maybe point out why? high complexity, large number of possible interleaving, humans cannot have good intuition about these things, etc.

%%
%% OUR OBJECTIVE
%%
Our main objective is thus to develop a \emph{systematic} and mathematically \emph{rigorous}, tool-supported specification and verification framework for concurrency control protocols and their isolation guarantees. 
Systematic means unified coverage of a wide range of isolation guarantees.
%in a unified manner. 
Rigor is guaranteed by mathematical proofs showing that \emph{all} possible database behaviors with arbitrarily many processes and transactions satisfy the desired isolation guarantee.

%%
%% WHAT WE DID TO ACHIEVE THIS GOAL
%% 
To achieve this goal, we developed \ourframework, the \emph{first mechanized} framework for \emph{rigorously} verifying isolation guarantees of transaction protocols.
\ourframework is based on the centralized
operational model of transactions proposed in~\cite{DBLP:conf/ecoop/XiongCRG19}, 
instantiatable by a range of prevalent isolation criteria, including SSER, SI, and TCC.
We formalize this model, called the \emph{abstract transaction model}, in Isabelle/HOL~\cite{DBLP:books/sp/NipkowPW02}, an interactive theorem prover for higher-order logic, which offers the expressiveness and strong soundness guarantees required for systematic and rigorous tool support.

\looseness=-1
To develop \ourframework, we also make several substantial enhancements to the original model, beyond its formalization, which underpin our protocol correctness proofs and their automation.
First, we significantly extend the abstract model with an extensive library of hundreds of lemmas, including those stating properties of the abstract model's underlying concepts (e.g., version lists and reader sets) and its invariants (e.g., read atomicity). The resulting \ourframework library is \emph{protocol-independent} and provides strong support for protocol correctness proofs.
Second, we set up Isabelle's automatic proof tools to apply many lemmas automatically, thereby providing considerable proof automation and reducing the user's proof burden.

\looseness=-1
Within \ourframework, we formalize transaction protocol designs as transition systems, \rev{where different types of transitions correspond to different protocol steps.} 
We establish their isolation guarantees by refinement of the appropriate instance of the abstract model, \rev{which ensures that the protocol's behaviors conform to the abstract model.}
We also identify a collection of generic protocol invariants that can support the refinement proofs of a wide range of protocols. \ourframework thus offers rigorous tool support for verifying isolation guarantees of new transaction protocols at an early design stage. %This mitigates the cost induced by redesign, implementation patching, and re-deployment of faulty designs.
% and avoids the cost of redesigning, patching, and redeploying faulty designs

To validate \ourframework, we analyze the isolation guarantees of two database designs. 
First, we model the classical Strict Two-Phase Locking (S2PL) protocol (combined with Two-Phase Commit) and we verify that it satisfies SSER.
With this case study, we illustrate different aspects of \ourframework including protocol specification, refinement mapping, and refinement proof using invariants.
Second, we show how to use \ourframework to systematically discover isolation bugs in database designs based on unprovable proof obligations, from which we construct concrete counterexamples.
As an illustrative example, we use the TAPIR~\cite{DBLP:journals/tocs/ZhangSSKP18} protocol, which claims to provide SSER. This protocol was previously found to exhibit a real-time ordering issue~\cite{osdi23}.
Using \ourframework, we discover a different counterexample, which violates \emph{atomic visibility}~\cite{ramp}, where a transaction's updates are only partially observed by others.

\inlsec{Contributions}
Overall, we make the following contributions.
\begin{itemize}[leftmargin=10pt]
    \item We develop \ourframework, the first mechanized framework for formally specifying  database designs and systematically verifying their isolation guarantees for \emph{all} their behaviors. Compared to existing work, \ourframework is mechanized, more expressive, covers more isolation levels, and offers  much stronger correctness guarantees.

    \item \rev{We model and verify the classical S2PL protocol combined with 2PC in \ourframework and prove that it satisfies SSER, thereby illustrating our protocol modeling and proof technique.}

    \item We demonstrate how \ourframework can be used to find isolation bugs in database designs by examining failed proof obligations.
    In particular, for TAPIR, we construct novel counterexamples illustrating its violation of atomic visibility. 
 
\end{itemize}
\section{Background}
\label{sec:background}

\subsection{Isolation Levels} \label{subsec:iso}
\begin{figure}[t]
\begin{center}
   \includegraphics[width=\columnwidth]{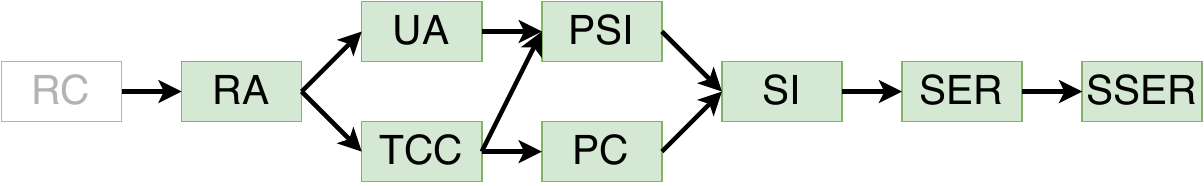}
\end{center}
   \caption{\looseness=-1 A hierarchy of prevalent 
   isolation levels.
    RC: read committed~\cite{si};
    RA: read atomicity~\cite{ramp};
    UA: update atomicity~\cite{ua,DBLP:conf/concur/Cerone0G15};
    TCC: transactional causal consistency~\cite{Cure:ICDCS2016,Eiger:NSDI2013};
    PC: prefix consistency~\cite{pc};
    SI: snapshot isolation~\cite{si}; 
    PSI: parallel SI~\cite{psi};
    SER: serializability~\cite{serializability}; 
    SSER: strict  SER~\cite{serializability}. $A \to B$ means $A$ is weaker than $B$.
   \ourframework covers the green levels.
  }
   \label{fig:isolation-hierarchy}
\end{figure}

Distributed databases provide various isolation levels, depending on the system's desired scalability and availability.
As shown in Figure~\ref{fig:isolation-hierarchy}, 
\ourframework supports well-known isolation levels such as SER and SI, as well as more recent guarantees such as RA~\cite{ramp} and TCC~\cite{Cure:ICDCS2016,Eiger:NSDI2013}.
We will briefly explain some of these isolation guarantees.

\begin{description}[leftmargin=10pt]
\item[Read Atomicity (RA).] This is also known as \emph{atomic visibility},
requiring that other transactions observe all or none of a transaction’s updates. 
It prohibits \emph{fractured reads}, \rev{such as in the scenario
\[
T_1\!: \mathtt{W}(A, \{B\}), \mathtt{W}(B, \{A\}) \qquad 
T_2\!: \mathtt{R}(A, \{B\}), \mathtt{R}(B, \emptyset),
\]}
\looseness=-1
where Carol \rev{(in transaction $T_2$)} only observes one direction of a new (bi-directional) friendship between Alice \rev{(A)} and Bob \rev{(B)} in a social network (recorded in $T_1$). \rev{RC allows such fractured reads as well as reading multiple versions of a key in one transaction.}

\item[Transactional Causal Consistency (TCC).] This level requires that two causally related transactions appear to all clients in the same causal order. 
It prevents \emph{causality violations}, \rev{such as
\[
\quad T_1\!: \mathtt{W}(A, m) \qquad T_2\!: \mathtt{R}(A, m), \mathtt{W}(B, r) \qquad T_3\!: \mathtt{R}(A, \bot), \mathtt{R}(B, r),
\]}
where Carol \rev{(in $T_3$)} observes Bob's response $r$ to Alice's message $m$ without seeing the message itself in a chatroom. \rev{Our notion of TCC also includes \emph{convergence}~\cite{Cure:ICDCS2016,Eiger:NSDI2013}, which requires different clients to observe all transactions in the same total order.}
In practice, most causally-consistent databases provide convergence.

\item[(Strict) Serializability ((S)SER).] \rev{SER requires that the effects of every 
concurrent execution can also be achieved by a sequential execution.} SSER additionally requires this sequential execution to preserve the real-time order of (non-overlapping) transactions.

\end{description}

\subsection{Operational Semantics for Transactions}
\label{subsec:background-operational-framework}

Xiong et al.~\cite{DBLP:conf/ecoop/XiongCRG19} introduced a state-based operational semantics for atomic transactions that operate on distributed key-value stores (KVSs). This semantics is formulated as a labeled transition system (\Cref{subsec:lts-refinement}), called the \emph{abstract transaction model}, which abstracts these KVSs into a single (centralized) multi-versioned KVS $\kvs\!: \keytype \fun \listtype{\verstype}$ that maps each key to a list of \emph{versions}. Each version~$\kvs(k,i)$ of a key~$k$ at list index $i$ records 
\begin{enumerate}[(i)]
%(i) 
\item the value stored, 
%(ii) 
\item the writer transaction, and 
%(iii) 
\item the reader set (the transactions that have read this version).
\end{enumerate}
The reader set tracks write-read (WR) dependencies. In a real, distributed system, each client $cl$ has a different partial \emph{client view} of~$\kvs$, and this is modeled by explicitly representing these views in the model as mappings $\views(cl) \!: \keytype \fun \powset{\nats}$, describing, for each key, the set of versions (denoted by list indices) visible to the client. 

The semantics assumes a \emph{last-write-wins} conflict resolution policy and the \emph{snapshot property}, which ensures that transactions read and write at most one version of each key.
%a client only sees one database snapshot. 
It also assumes that views are \emph{atomic}, i.e., clients observe either all or none of a transaction's effects. These properties together ensure \emph{atomic visibility} and establish RA as the model's baseline isolation guarantee.
Transactions are described by a \emph{fingerprint} $\fprint \! : \keytype \times \rwtype \map \valuetype$, which maps each key and operation (read or write) to a value, if any.

\looseness=-1
The model has two types of transitions: \emph{atomic commit}, which atomically executes and commits an entire transaction,  
and \emph{view extension}, which monotonically extends a client's view of the KVS.
A commit transition can only be executed under certain conditions on the current KVS~$\kvs$,  the client's view $\views(cl)$, and the transaction's fingerprint $\fprint$. These conditions depend on the desired isolation guarantee. The framework is parameterized on these conditions and can be instantiated to eight different isolation guarantees (Figure~\ref{fig:isolation-hierarchy}). 

% salient features:

% - global, centralized multi-versioned KVS $\kvs\!: \keytype \fun \verstype$. 

% - versions include value, writer transaction, and reader set

% - (partial) client views $u \!: \keytype \fun \powset{\nats}$

% - execution tests: determine whether a client with a given view can commit a transaction

% - last-write-wins and snapshot property yield RA baseline model

% - other supported isolation levels, see Figure~\ref{fig:isolation-hierarchy}

% - can be used to prove that (i) protocols satisfy desired isolation guarantees and (ii) client programs satisfy certain invariants under these isolation levels; this ability is new compared to other levels (most declarative and a few operational ones)

\subsection{\rev{Isabelle/HOL and Notation}}
\label{subsec:isabelle-hol}

\looseness=-1
\rev{We use the Isabelle/HOL proof assistant for our modeling and proofs~\cite{DBLP:books/sp/NipkowPW02,nipkow2014concrete}. Isabelle is a generic framework for implementing logics, and Isabelle/HOL is its instance to higher-order logic (HOL). HOL can be paraphrased as ``functional programming plus quantifiers''. 
Isabelle offers powerful automated reasoning tools, including an integration of various external (first-order logic) automated theorem provers. 
We chose Isabelle/HOL for its high expressiveness and good proof automation, which allows us to naturally formalize the framework's structures and obtain strong protocol correctness guarantees for arbitrary numbers of processes and transactions.}

\subsubsection*{Notation}
To enhance readability, we use standard mathematical notation where possible and blur the distinction between types and sets. 
For a function $f:A\fun B$, we denote by $f[x\mapsto y]$ the function that maps $x$ to $y$ and otherwise behaves like $f$. For a partial function $g: A \map B$, we write $dom(g)$ for the domain of $g$, and $g(x) = \bot$ if $x \notin dom(g)$. \rev{We also define $A_{\bot} = A \uplus \{\bot\}$.} For a relation $R \subseteq A \times A$, $R^{-1}$ denotes its converse and $R^{+}$ its transitive closure. 

\section{\rev{System Modeling and Refinement}}
\label{subsec:lts-refinement}

\rev{We introduce the modeling and verification formalism~\cite{AL91,DBLP:journals/iandc/LynchV95} we use in our work and illustrate it with a simple database example.}

\subsection{System Models}

We use \emph{labeled transition systems} (LTS) to model database protocols and the abstract transaction model. An LTS $\mathcal{E} = (S, I, \{\myrightarrow{e} \, \mid e \in E\})$ consists of a set of states $S$, a non-empty set of initial states $I \subseteq S$, and a family of transition relations $\myrightarrow{e} \;\subseteq S \times S$, one for each event $e \in E$. We assume that our LTS models contain an idling event $\Skip$, defined by $s \myrightarrow{\Skip} s$, which we use in refinements (see below).
We often define the relations $\myrightarrow{e}$ using \emph{guard} predicates $G_e$ and \emph{update} functions $U_e$ by $s \myrightarrow{e} s'$ if and only if $G_e(s) \land s' = U_e(s)$. 
\begin{example}[\rev{Atomic Transactions}]
\label{ex:simple-db}
\rev{We model a simple centralized database system with a single-version KVS and atomic (one-shot) transactions as an LTS $\mathcal{A}$. 
A state of our LTS  $\mathcal{A}$ is a KVS, which we represent as a function $\mathcal{S}: \keytype \fun \valuetype$, from keys to values. All states are initial states. 
Besides $\Skip$, the LTS  $\mathcal{A}$ has a single (parametrized) event $\mathsf{txn}(f_R,f_W)$, where the parameters $f_R,f_W:\keytype \map \valuetype$ respectively denote a read and a write fingerprint. We specify the associated transition relation as 
\[
  \inlisa{txn}(f_R,f_W) : \; 
  f_R \subseteq \mathcal{S} \:\land\: \mathcal{S}' = \mathcal{S} \triangleright f_W,
\]
meaning that $\mathcal{S} \xrightarrow{\inlisa{txn}(f_R,f_W)} \mathcal{S}'$ is defined by the predicate after the colon. Here, $f_R \subseteq \mathcal{S}$ is the guard, requiring that the values read are those in the KVS $\mathcal{S}$, and $\mathcal{S}' = \mathcal{S} \triangleright f_W$ is the update, stating that the KVS $\mathcal{S}'$ after the transition equals the KVS $\mathcal{S}$ updated with the values in $f_W$. The updated KVS $(\mathcal{S} \triangleright f_W)(k)$ is defined by $f_W(k)$ if $k \in dom(f_W)$ and $\mathcal{S}(k)$ otherwise. Note that the parameter $f_R$ can be considered as an output (the values read) and $f_W$ as an input (the values to be written) to the event. Hence, in this very abstract model $\mathcal{A}$, all of a transaction's reads and writes are executed in a single atomic state transition. }
\end{example}

\subsection{Invariants and Refinement}

\looseness=-1 \rev{We define invariants as supersets of the set of reachable states (or, equivalently, predicates holding on all reachable states), and refinement as a process for successively adding features to models.}

A state $s$ is \emph{reachable} if there is a sequence of transitions from an initial state leading to $s$. We denote the set of reachable states of $\mathcal{E}$ by $\reach(\mathcal{E})$. A set of states $J$ is an \emph{invariant} if $\reach(\mathcal{E}) \subseteq J$. Invariants are proved by showing that they hold in all initial states and are preserved by all state transitions.

\emph{Refinement} relates two LTSs 
$\mathcal{E}_i = (S_i, I_i, \{\myrightarrow{e}_i \, \mid e\in E_i\})$, for $i\in\{1,2\}$,
%$\mathcal{E}_1 = (S_1, I_1, \{\myrightarrow{e}_1 \, \mid e\in E_1\})$ and 
%$\mathcal{E}_2 = (S_2, I_2\, \{\myrightarrow{e}_2\, \mid e \in E_2\})$, 
which usually represent different abstraction levels. 
Given \emph{refinement mappings} $r\!: S_2 \fun S_1$ and $\pi\! : E_2 \fun E_1$ between the LTSs' states and events, we say $\mathcal{E}_2$ refines $\mathcal{E}_1$, written $\mathcal{E}_2 \refines_{r,\pi} \mathcal{E}_1$, \rev{if for all states $s,s' \in S_2$ and events $e \in E_2$,}
\begin{enumerate}[(i)]
\item $r(s) \in I_1$ whenever $s \in I_2$, and 
\item $r(s)$ \raisebox{-1pt}{$\myrightarrow{\pi(e)}_1$} $r(s')$ whenever $s \myrightarrow{e}_2 s'$. 
\end{enumerate}
\rev{We can assume without loss of generality that the state $s$ in these proof obligations is reachable.
This allows us to use invariants of the concrete model in refinement proofs, which is often necessary.}
When using guards and updates, and an invariant $J$ of $\mathcal{E}_2$, (ii) reduces to two proof obligations: assuming $G^2_e(s)$ and $s \in J$, prove 
\begin{enumerate}[(a)]
%(a) 
\item $G^1_{\pi(e)}(r(s))$ (\emph{guard strengthening}), and 
%(b) 
\item $r(U^2_e(s)) = U^1_{\pi(e)}(r(s))$ (\emph{update correspondence}).
\end{enumerate}
\rev{Guard strengthening ensures that whenever the concrete event is executable then so is the corresponding abstract event. Update correspondence expresses that the refinement mapping $r$ commutes with the concrete and abstract state updates.}

Refinement guarantees the inclusion of sets of reachable states (modulo~$r$), i.e., $r(\reach(\mathcal{E}_2)) \subseteq \reach(\mathcal{E}_1)$. 
\rev{Hence, $\mathcal{E}_1$ specifies which states $\mathcal{E}_2$ is allowed to reach (modulo $r$).}

\begin{example}[\rev{Refined transactions}]
\rev{We define a more concrete model $\mathcal{C}$. Instead of executing transactions in a single atomic transition, we have three operations, read, write, and commit, which we model as separate events.
This model's states consist of triples $(\mathcal{S}, \rfprint, \wfprint)$, where $\mathcal{S}$ is as before, and $\rfprint$ and $\wfprint$ are read and write fingerprints, which are now part of the state. Initial states are those with empty fingerprints.
The model $\mathcal{C}$'s three events and their transition relations are as follows:
\begin{align*}
  \inlisa{read}(k, v)
  & : \;
  \mathcal{S}(k) = v \,\land\, \rfprint' = \rfprint[k \mapsto v] 
\\
  \inlisa{write}(k, v)
  & : \;
  \wfprint' = \wfprint[k \mapsto v] 
\\
  \inlisa{commit}(f_R,f_W) 
  & : \;
  f_R = \rfprint \,\land\, f_W = \wfprint \,\land\, 
\\
  & \phantom{: \;\;\;}
  \mathcal{S}' = \mathcal{S} \triangleright \wfprint \,\land\, 
  \rfprint' = \emptyset \,\land\, \wfprint' = \emptyset
\end{align*}
The read event's guard checks that the KVS $\mathcal{S}$ maps the key $k$ to the value $v$ and updates the read fingerprint $\rfprint$ with this mapping. The write event updates the write fingerprint $\wfprint$ with the given key-value mapping. 
Finally, the commit event's guard just binds the current fingerprints to the respective event parameters. This event updates the KVS $\mathcal{S}$ with the write fingerprint $\wfprint$ as in \Cref{ex:simple-db} and resets both fingerprints to the empty maps. Hence, this model only allows the \emph{serialized} execution of transactions, one at a time. }

\end{example}

\begin{example}[\rev{Refinement proof}]
\rev{We now prove $\mathcal{C} \refines_{r,\pi} \mathcal{A}$ for suitable refinement mappings $r$ and $\pi$. We define the state mapping $r$ by $r(\mathcal{S}, \rfprint, \wfprint) = \mathcal{S}$ and the event mapping $\pi$ as $\pi(\inlisa{commit}(f_R,f_W)) = \inlisa{txn}(f_R,f_W)$, with all other events mapping to $\Skip$. This means that only the commit event has an observable effect on the abstract state and this effect corresponds to an abstract transaction.}

\rev{For the refinement proof, note that the proof obligations for the read and write events trivially hold as these events do not change $\mathcal{S}$. The interesting case is the commit event. It is easy to see that update correspondence holds, since the concrete commit and the abstract $\inlisa{txn}$ events' KVS updates are identical. For the guard strengthening proof obligation, we must show that $f_R = \rfprint \,\land\, f_W = \wfprint$ implies $f_R \subseteq \mathcal{S}$. This is not possible, as the concrete guard provides no information about the KVS $\mathcal{S}$. 
However, inspecting the concrete system $\mathcal{C}$, we see that the read event only records key-value mappings in $\rfprint$ that are also present in $\mathcal{S}$. Hence, we can establish $\rfprint \subseteq \mathcal{S}$ as invariant of $\mathcal{C}$. We can then use this invariant together with the guard $f_R = \rfprint$ to complete the refinement proof.}
\end{example}

\section{The \ourframework Formal Framework} %and Application}
\label{sec:formal-framework-and-application}

\rev{Our framework is centered around our formalization of the abstract transaction model introduced in \Cref{subsec:background-operational-framework}.}
We formalize the abstract model in Isabelle/HOL as an LTS $\isomodel{\isolevel}$, parameterized by an isolation level $\isolevel \in \{\RA, \UA, \TCC, \PSI, \CP, \SI, \SER, \SSER\}$ (cf.~\Cref{fig:isolation-hierarchy}). \rev{We describe this formalization in \Cref{subsec:abstract-model}.}

\rev{Given a pseudocode specification of a concurrency control protocol design, its modeling and verification using our framework consists of the following three steps:}

\begin{enumerate}
\item \rev{Formalize the pseudocode specification of the protocol as an LTS model $\mathcal{M}$ in Isabelle/HOL, as discussed in \Cref{subsec:protocol-modeling}. }

\item \rev{Specify the protocol's desired isolation level $\isolevel$ by the appropriate instance, $\isomodel{\isolevel}$, of the abstract transaction model.}

\item \rev{Define refinement mappings on states $r$ and events $\pi$ and prove that $\mathcal{M}$ refines the abstract transaction model instance $\isomodel{\isolevel}$.
%, i.e., $\mathcal{M} \refines_{r,\pi} \isomodel{\isolevel}$. 
This is explained in \Cref{subsec:protocol-verification-technique}.}
\end{enumerate}

\subsection{Formalizing the Abstract Transaction Model}
\label{subsec:abstract-model}

\begin{figure}[t]
\begin{lstlisting}
datatype txid = Tn(nat, cl_id)

record version =
  v_value : value
  v_writer : txid
  v_readerset : set(txid)

type_synonym kv_store = key $\fun$ list(version)
type_synonym view = key $\fun$ set(nat)
type_synonym config = kv_store $\times$ (cl_id $\fun$ view) 
\end{lstlisting}
\captionsetup{skip=3pt}
\caption{\rev{Configurations of abstract transaction model.}}
\label{fig:abs-model-configs}
\end{figure}

\subsubsection{\rev{Parametrized} abstract transaction model}
\looseness=-1 The model's states $(\kvs, \views)$, called configurations, consist of a central KVS $\kvs$ and the client views $\views$, \rev{mapping client IDs to their views, as explained in \Cref{subsec:background-operational-framework}. This structure is formalized in \Cref{fig:abs-model-configs}. In particular, we model transaction identifiers as a datatype \inlisa{txid}, whose elements $\txid{sn}{cl}$ are indexed by the issuing client $cl$ and a 
%(monotonically increasing)
sequence number $sn$, and we model versions as a record type \inlisa{version} in Isabelle/HOL.}

The model has three events: an (atomic) commit event, a view extension event, and the idling event.
%(cf.~\Cref{subsec:background-operational-framework}).
%We now give a high-level description of the set of guards that execution tests must satisfy to be executable and the resulting configuration update. 
The commit event's guard depends on $\isolevel$, which is specified by two elements:
\begin{itemize}[leftmargin=15pt]
\item a relation $R_{\isolevel} \subseteq \inlisa{txid} \times \inlisa{txid}$ on transaction identifiers, and 
\item a predicate $\vShift_{\isolevel}(\kvs, u, \kvs', u')$ on two KVSs and two views. 
\end{itemize}
We will further describe $R_{\isolevel}$ and $\vShift_{\isolevel}$ below.

%\begin{definition}[Abstract commit] \label{def:et_trans}
The abstract commit event's transition relation 
\rev{\[
(\kvs, \views) \xrightarrow{\isaco{commit}(cl, \isa{sn}, u, \fprint)}_{\isolevel} 
  (\kvs', \views')
\]
is defined in Isabelle/HOL by: }
\smallskip
\begin{lstlisting}
definition commit($cl,sn,u,\fprint,(\kvs,\views),(\kvs',\views')$) $\longleftrightarrow$
  $\views(cl) \sqsubseteq u$ $\land$ wf($\kvs,u$) $\land$ wf($\kvs',u'$) $\land$   -- basic view guards
  LWW($\kvs,u,\fprint$) $\land\,\,$                        -- last-write-wins
  Tn($sn,cl$) $\in$ nextTxids($\kvs,cl$) $\land$         -- txid freshness
  canCommit($\kvs,u,\fprint,R_{\isolevel}$)$\;\land\;$vShift$_{\isolevel} (\kvs,u,\kvs',u'$) $\land$ --$\mbox{\color{olive}{\,\isolevel}}$-specific
  $$$\kvs$'$=\:$UpdateKV($\kvs,\,$Tn($sn,cl$)$,u,\fprint$)$\;\land\;\views$'$=\views[cl\mapsto\,u']$ --$\,$updates
\end{lstlisting}
\smallskip

The transition updates the configuration $(\kvs, \views)$ to the new configuration $(\kvs', \views')$, where $\kvs'$ is the updated KVS and $\views'$ updates the client $cl$'s view  to $u'$. In particular, $\kvs'$ is obtained from $\kvs$ by recording the operations described by the fingerprint~$\fprint$, i.e., the writes append a new version with writer ID $\txid{sn}{cl}$ to the respective key's version list, and the reads add $\txid{sn}{cl}$ to the respective versions' reader sets. 
The guards have the following meanings:
\begin{itemize}[leftmargin=15pt]
\item $\views(cl) \sqsubseteq u$ allows one to extend the client $\isa{cl}$'s current view to a (point-wise) larger one before committing. 

\item \looseness=-2 $\wellformed(\kvs, u)$ and $\wellformed(\kvs', u')$ require $u$ and $u'$ to be \emph{wellformed views}, i.e., atomic and holding indices that point to existing versions. %all pointed-to versions exist.

\item \looseness=-1 \rev{$\LWW(\kvs, u, \fprint)$ captures the \emph{last-write-wins} conflict resolution policy, where a client reads each key's latest version in its view~$u$. If $\fprint(k, \inlisa{R}) = v$ then $v$ must be the version $\kvs(k, i)$'s value, where $i$ is the highest index in $u(k)$.} 

\item $\txid{sn}{cl} \in \isaco{nextTxids}(\kvs, cl)$ represents the transaction ID freshness requirement, i.e., the sequence number $sn$ is larger than any of the client $cl$'s sequence numbers used in~$\kvs$.

\item $\canCommit(\kvs, u, \fprint, R_{\isolevel})$ is the central commit condition, which ensures that it is safe to commit a transaction at a given isolation level $\isolevel$. It requires that the set of visible transactions $\visTx(\kvs,u)$ (i.e., the writers of the versions that the view $u$ points to) is \emph{closed} under the relation $R_{\isolevel}$ in the sense that 
\begin{equation}
\label{eq:closedness} %\tag{$\mathcal{C}(\isolevel)$}
  (R_{\isolevel}^{-1})^{+}(\visTx(\kvs,u)) \subseteq \visTx(\kvs,u) \cup \readonly(\kvs).  
\end{equation}
In other words, following the relation $R_{\isolevel}$ backwards from visible transactions, we only see visible or read-only transactions.

\item \looseness=-1 $\vShift_{\isolevel}(\kvs, u, \kvs', u')$ constrains the allowed modifications of the client view during the commit. This predicate captures session guarantees like monotonic reads and read-your-writes.   

\end{itemize}

We define the view extension event's transition relation
\rev{\[
(\kvs, \views) \xrightarrow{\isaco{xview}(cl, u)}_{\isolevel} 
  (\kvs', \views')
\]
in Isabelle/HOL by:} 
\begin{lstlisting}
definition xview(cl,$\,$u,$\,$($\kvs$,$\,\views$),$\,$($\kvs$',$\,\views$')) $\longleftrightarrow$
  $\views$(cl) $\sqsubseteq$ u $\land$ wf($\kvs$,$\,$u) $\land$
  $\kvs$' = $\kvs$ $\land$ $\views$' = $\views$[cl $\mapsto$ u]
\end{lstlisting}

This event simply extends a client $cl$'s view from $\views(cl)$ to a wellformed view $u$. It abstractly models that additional versions of certain keys become visible to the client.

\subsubsection{Instantiation to concrete isolation levels}
To instantiate this abstract model for a particular isolation level $\isolevel$, one specifies the relation $R_{\isolevel}$ and the predicate $\vShift_{\isolevel}$. We describe here the definitions for $\isolevel \in \{\rev{\RA,} \TCC, \SSER\}$ and we refer the reader to~\cite{DBLP:conf/ecoop/XiongCRG19} for other instantiations. 
\rev{For the model's baseline isolation level $\RA$, the conditions $\canCommit$ (with $R_{\RA} = \emptyset$) and $\vShift_{\RA}$ always hold.}
For $\TCC$, we define $R_{\TCC} = \SO \cup \WR_{\kvs}$, where $\SO$ captures the clients' session orders and $\WR_{\kvs}$ is the write-read dependency between transactions in~$\kvs$, determined by the reader set associated to each version. 
The resulting commit condition $\canCommit_{\TCC}$ requires that the views are closed under \emph{causal dependencies}. The predicate $\vShift_{\TCC}$ captures the monotonic reads and read-your-writes session guarantees. 
For $\SSER$, we have $R_{\SSER} = \WW_{\kvs}^{-1}$, where $\WW_{\kvs}$ is the write-write dependency (i.e., per-key version order) on transactions in~$\kvs$. Hence, the commit condition $\canCommit_{\SSER}$ expresses that the views must include \emph{all} versions in $\kvs$ before the commit. The condition $\vShift_{\SSER}$ is true, i.e., it always holds.

\rev{Note that the semantics' built-in atomic visibility 
%and convergence (implied by the totally ordered version lists) 
precludes the representation of the 
%TCC and 
RC isolation level (cf.~\Cref{fig:isolation-hierarchy}).} 
In future work, we will extend the model to also cover this level.

\subsubsection{Discussion of formalization}
\looseness=-1
The formalization involved both design choices and adaptations compared to ~\cite{DBLP:conf/ecoop/XiongCRG19}. 
First, to represent the framework's elements in Isabelle/HOL, we used Isabelle's type system to avoid proving invariants where possible. For example, we represent fingerprints as partial functions rather than relations.
Second, the $\canCommit$ condition in~\Cref{eq:closedness} is an equivalent reformulation of the original one. We find our version easier to understand and prove. \ourframework's library provides proof rules that help establish this condition.
Third, the paper~\cite{DBLP:conf/ecoop/XiongCRG19} defines a KVS wellformedness condition  including the snapshot property, and \emph{assumes} it for all KVSs. In our formalization, we \emph{proved} KVS wellformedness as an invariant of the abstract model.

\subsection{Protocol Modeling} 
\label{subsec:protocol-modeling}

\begin{figure}[t]
\begin{lstlisting}[numbers=none]
function cl_compute(tasks):             -- client
  for each server svr:
    results(svr) = svr_compute(tasks(svr), svr)
  return aggregate(results)

function svr_compute(task, svr):        -- servers 
  result = compute(task)
  return result
\end{lstlisting}
\captionsetup{skip=0pt}
\caption{\rev{Pseudocode for remote computation.}}
\label{fig:example-pseudocode}
\vspace{-.6ex}
\end{figure}

\rev{We assume that protocol designs are given as pseudocode specifications for clients and servers. We explain how to transform such specifications into a formal LTS protocol model in Isabelle/HOL.} 

\rev{For illustration, we use a simple remote computation example, where clients outsource computation tasks to servers using (asynchronous) remote procedure calls (RPCs) and then collect and aggregate the servers' results into a final result (\Cref{fig:example-pseudocode}). The same modeling technique also applies to concurrency control protocols. We discuss some points specific to those protocols at the end of this subsection.}
Formalizing a protocol design as an LTS in Isabelle/HOL requires the definition of the LTS' global configurations and events.

\subsubsection{\rev{Protocol configurations}}

\begin{figure}[t]
\begin{lstlisting}
datatype cl_state =     -- client state
  cl_idle | cl_invoked(tasks) | cl_done(final)

record cl_conf =        -- client configuration
  cl_state : cl_state

datatype comp_state =   -- computation state
  svr_idle | svr_done(result)

record svr_conf =       -- server configuration
  svr_state : cl_id $\fun$ comp_state

record global_conf =    -- global configuration
  cls : cl_id $\fun$ cl_conf
  svrs : svr_id $\fun$ svr_conf
\end{lstlisting}
\captionsetup{skip=0pt}
\caption{\rev{Client, server, and global configurations.}}
\label{fig:example-configs}
\vspace{-1ex}
\end{figure}

\looseness=-1
Protocol configurations are composed of the clients' and the servers' local configurations \rev{(\Cref{fig:example-configs}). 
A client's configuration consists of the client's control state (of type \inlisa{cl\_state}), which here may be idle, invoked with given tasks, or done with the final computation result.
The server has a computation state (of type \inlisa{comp\_state}) per request. We assume for simplicity that each client only invokes a server once, so we can identify requests with clients. A computation's state is either idle or done with a result.}

\subsubsection*{Notation.} 
Given a global configuration \inlisa{s}, the client \inlisa{cl}'s state, for instance, is accessed by \inlisa{cl\_state(cls(s,cl))}, where the functions
\inlisa{cls} 
and \inlisa{cl\_state} 
are record field projections.
For readability, we will write this as \inlisa{cl\_state(cl)} when \inlisa{s} is clear from the context. We will use \inlisa{cl\_state'(cl)} for the field's value in the successor state \inlisa{s'} of an event. Similar shorthands apply to all record fields.

\subsubsection{\rev{Protocol events}}

\begin{figure}[t]
\begin{lstlisting}
-- client-side events
definition cl_compute_invoke($cl,\isaid{tasks}$) $\longleftrightarrow$
  cl_state($cl$) = cl_idle $\land$
  cl_state'($cl$) = cl_invoked($\isaid{tasks}$)

definition cl_compute_response($cl,\isaid{results},\isaid{final}$) $\longleftrightarrow$
  cl_state($cl$) = cl_invoked(_) $\land$
  ($\forall \isaid{svr}.$ svr_state($\isaid{svr},cl$) = svr_done(results($\isaid{svr}$)) $\land$
  $\isaid{final}$ = aggregate($\isaid{results}$) $\land$
  cl_state'($cl$) = cl_done($\isaid{final}$)

-- server-side event
definition svr_compute($svr,cl,task$) $\longleftrightarrow$
  svr_state($svr,cl$) = svr_idle $\land$
  ($\exists tasks.$ cl_state($cl$) = cl_invoked($tasks$) $\land$ 
           $task$ = tasks($svr$)) $\land$ 
  svr_state'($svr,cl$) = svr_done(compute($task$))
\end{lstlisting}
\captionsetup{skip=0pt}
\caption{\rev{Events for remote computation example.}}
\label{fig:example-events}
\vspace{-1.5ex}
\end{figure}

\looseness=-1
\rev{We model protocol events according to the following principles.} 
Events only change either a single client or a single server's configuration. This ensures that clients and servers can be seen as independent components with interleaved events. 
To communicate, these components directly access each other's configurations. \rev{More precisely, the destination (or receiver) component reads the desired information from the data owner (or sender). The sender just provides the information, but does not actively communicate itself (e.g., send a message).}
This is a standard abstraction in protocol modeling, which can later be refined into explicit message passing through a channel. 
%communication medium. 
%
\rev{Events may have parameters, which correspond to the free variables in the event specification.}

\rev{We translate the protocol's pseudocode from \Cref{fig:example-pseudocode} into two client events, one for the %\new{asynchronous} 
RPC request and the other for the response, and one server event, which handles the entire RPC on the server side (\Cref{fig:example-events}). 
The client event \inlisa{cl\_compute\_invoke} transits from idle to the invoked state, storing the given \inlisa{tasks}. Note that no client-server communication is modeled in this event.
The server event \inlisa{svr\_compute} reads its task $\inlisa{tasks(svr)}$ from client \inlisa{cl}'s invoked state, which models receiving a message containing  \inlisa{tasks(svr)} from \inlisa{cl}. The existential quantifier on Line~15 means that the \inlisa{tasks} map itself is hidden from the server, i.e., not transmitted.
Finally, the client event \inlisa{cl\_compute\_response} starts in an invoked state, where the tasks stored do not matter, as indicated by the underscore (Line~7), collects the results from all servers into the \inlisa{results} map (Line~8), computes the \inlisa{final} result (Line~9), and stores it in its new done state (Line~10).
Note that record fields not mentioned in an event's update remain unchanged.}

\subsubsection{\rev{Modeling concurrency control protocols}}

We consider distributed systems with per-client transaction coordinators and several shard/replica servers that handle the clients' reads and writes to keys. 
For simplicity, we integrate the coordinator into the client and assume that each server stores one key.
We focus on concurrency control and do not model (orthogonal) replication aspects.

\looseness=-1
\rev{Most concurrency control protocols follow a two-phase commit structure, which is reflected in the client's and the server's states. We assume the clients perform transactions sequentially, whereas the servers handle an arbitrary number of concurrent transactions. Hence, the servers will have per-transaction states. Both client and server configurations may contain additional fields to store protocol-specific information such as clocks, timestamps, or version lists.}

\subsection{Protocol Verification}
\label{subsec:protocol-verification-technique}

\looseness=-1
\rev{After modeling a concurrency control protocol, we prove by refinement that it implements the desired instance of the abstract transaction model. We show how to define refinement mappings relating the protocol model to the abstract one and conduct the refinement proof. 
We also briefly discuss an advanced proof technique for the case when refinement alone is insufficient to prove correctness. }

\subsubsection{\rev{Defining the refinement mappings}} 
\label{subsubsec:refinement-mappings}

\rev{To relate the protocol and abstract models, we must define the refinement mappings~$r$ on states and~$\pi$ on events. 
To define $\pi$, we identify one or more protocol events that correspond to the protocol's transaction commit points and we map these events to the abstract commit event. These are usually client-side (i.e., coordinator) commit events. Events that change a client's view of the KVS (e.g., by making additional versions visible) refine the abstract view extension event. All remaining events must refine the abstract idling event ($\Skip$), meaning they have no corresponding effect in the abstract model.}

To define~$r$, we reconstruct an abstract configuration $r(s)$ %$(\kvs, \views)$ 
from a concrete protocol configuration $s$, i.e.,
\[
  \refmap{r}{}(s) = (\refmapkvs{}(s), \refmapviews{}(s)).
\]

The function $\refmapkvs{}(s)$ reconstructs each key's version list from the corresponding server's state and extends it with the effects of all ongoing transactions' \emph{client-committed} operations, i.e., those operations where the client has already committed in the two-phase commit protocol, but the corresponding server has not yet followed to their committed state. 
%Could drop next sentence if space needed.
In particular, the client-committed reads are added to the appropriate reader sets and the writes are appended as new versions to the respective key's version list. 
This part of $r$ is largely determined by the need to satisfy update correspondence, i.e., that abstract and concrete state updates commute with $r$.

The function $\refmapviews{}(s)$ reconstructs the abstract configuration's client views from the protocol configurations. 
\rev{For example, in some protocols for achieving TCC, clients maintain an explicit threshold (e.g., a timestamp) for reading ``safe'' versions, which are guaranteed to be committed. In this case, the view would include all versions below that threshold. In other protocols (e.g., for achieving SSER), clients must read the latest version of each key. Here, one can use full views of each key, which include all versions.}
This part of $r$ is mostly constrained by the need to satisfy the view-related guards of the abstract commit event (wellformedness, \LWW, \canCommit, \vShift) when refining the protocol's committing events.

To help define~$r$, we often add \emph{history variables} to the protocol configurations. These record additional information, such as the order of client commits for each key to reconstruct the abstract version lists. However, they are not required for the protocol execution itself. In particular, event guards must not depend on them.

\subsubsection{Refinement proofs} 
\label{subsubsec:refinement-proofs}

To establish the desired refinement 
\begin{equation}
\label{eq:protocol-refinement}
\mathcal{M} \refines_{r,\pi} \isomodel{\isolevel},
\end{equation}
we must prove for each concrete event 
(i) guard strengthening, i.e., that the protocol event's guards imply the \rev{abstract event's guards},
%(see \Cref{subsec:abstract-model}),
and 
(ii) update correspondence, i.e., that \rev{the abstract and concrete updates commute with the refinement mapping.}
%abstracting the result of the protocol configuration's concrete update equals the result of the abstract update of the abstracted protocol configuration.
%(cf.~\Cref{subsec:lts-refinement})
\ourframework provides proof rules that support such refinements. 
Recall that the refinement~\eqref{eq:protocol-refinement} implies the following \emph{correctness condition}: 
\begin{equation}
\label{eq:reach-inclusion}
r(\reach(\mathcal{M})) \subseteq \reach(\isomodel{\isolevel}),
\end{equation}
where $r$ is the state component of the refinement mapping.

\rev{We typically approach such proofs in the following way. We first prove the easy cases of concrete events that map to the abstract $\Skip$ event. For these, guard strengthening is trivial and update correspondence shows that the concrete event has no effect on the abstract state. 
More interesting are the events refining the abstract commit event (and also view extension). Here, update correspondence ensures that the protocol's concrete KVS updates are consistent with the abstract centralized KVS. Most of the work goes into proving guard strengthening, i.e., establishing the abstract commit's guards from the concrete event's guards. 
By first proving the basic guards, i.e., view extension ($\views(cl) \sqsubseteq u$) and view wellformedness, followed by $\LWW$, we know that the protocol satisfies the baseline RA isolation level. We then establish the guards $\canCommit$ and $\vShift$, which are specific to the desired isolation level.}

\rev{In case we are unable to establish a proof obligation, most often an abstract commit guard, this may happen for different reasons: 
\begin{enumerate}[leftmargin=2em, label={\arabic*.}]
%First, 
\item We may discover a \emph{mistake} in the protocol modeling or in the definition of the refinement mapping, which we must correct. 
% Second, 
\item We may \emph{miss some information} to complete the proof. In this case, we try to prove a protocol \emph{invariant} providing that information. For example, Isabelle's proof state may indicate a prepared client and a corresponding committed server, which is a contradiction derivable from a suitable invariant relating client and server states. Below, we list some typical invariants.
% Third, 
\item If neither of the previous two cases applies, further analysis may indicate a \emph{counterexample} violating the desired isolation level. We will explain this case in \Cref{sec:tapir}. Note that a counterexample may also arise from a failed attempt to prove a required protocol invariant.
\end{enumerate}
This is an iterative process that continues until all proof obligations are established or a counterexample is found.}

\subsubsection{\rev{Protocol invariants}}
\label{subsubsec:protocol-invariants}
Refinement proofs invariably require different protocol invariants, many of which are recurring and thus reusable for other protocols. The most important ones concern:
\begin{description}[leftmargin=10pt]
\item[Freshness of transaction IDs] The clients' current transaction ID is fresh, i.e., does not occur in the KVS until the commit;
\item[Past and future transactions] stating that the respective client and servers are in particular starting or end states; and
\item[Views] These invariants include view wellformedness, view closedness (for \canCommit), and session guarantees (for \vShift).
\end{description}
Many of these invariants directly imply related guards needed in the refinement of the abstract commit event. 
Moreover, there are invariants related to particular protocol mechanisms such as locking (cf.~\Cref{sec:two-phase-locking}) or timestamps.

\subsubsection{Advanced verification technique}

Sometimes, refinement is insufficient to establish the correctness condition in \Cref{eq:reach-inclusion}.
In particular, protocols relying on \emph{timestamps} to isolate transactions, such as those commonly seen for \emph{optimistic} concurrency control or for achieving TCC isolation, usually use the timestamps to define an order on versions and to identify ``safe-to-read'' versions (defining their view).
Hence, to ensure that clients always read the latest version in their view, the refinement mapping must reconstruct the version lists of the abstract KVS in the order of their commit timestamps. 
However, for such protocols, the \emph{execution order} of commits and the \emph{order of the associated commit timestamps} may differ and executions with such \emph{\invcmts} may thus require \emph{inserting} rather than \emph{appending} a key's new version to its version list. Since the abstract model always appends new versions to the version lists, a refinement proof alone would fail for such executions. 

\looseness=-1
Proving the correctness condition in \Cref{eq:reach-inclusion} for such protocols therefore requires the use of an advanced proof technique based on Lipton's reduction method~\cite{DBLP:journals/cacm/Lipton75} to reorder and eliminate inverted commits prior to the refinement proof. \ourframework also provides proof rules for such reduction proofs. \rev{In~\cite{GhasemiradSprengerLiu+-TACAS25}, we explain this technique in more detail and we apply it to a protocol achieving TCC.} 

% This is a proof rule:
% \[
% \infer[name]{C}{
%   A & B & D & E & F & G& H& I & J
% }
% \]

% \begin{align*}
%    bla & bla & bla & bla\\
%    x & y & z & u
% \end{align*}

\section{Strict Two-phase Locking}
\label{sec:two-phase-locking}

\begin{figure}[t]
\begin{lstlisting}[numbers=none]
-- client-side (transaction manager)
function initiateTransaction(transaction):
	for each RM in transaction.participants:   -- 2PC: prepare
		resp[RM] = prepare(transaction, RM)
	$\mbox{\color{violet}{\texttt{\textbf{if}}}}$ all participants respond with "okay":
		for each RM in transaction.participants: -- 2PC: commit
	    	commit(transaction, RM)
	else: -- abort if any RM responds "not okay"
		for each RM in transaction.participants:
	    	abort(transaction, RM)

-- server-side (resource manager)
function prepare(transaction, RM):
	try:
		acquireLocks(transaction.resources[RM])  -- S2PL: grow
    performOperations(transaction.operations[RM])
    return "okay"
	except:
		return "not okay"

function commit(transaction, RM):
	commitTransaction(transaction)       -- finalize changes
	releaseLocks(transaction.resources[RM])  -- S2PL: shrink

function abort(transaction, RM):
	abortTransaction(transaction)        -- rollback changes
	releaseLocks(transaction.resources[RM])  -- S2PL: shrink
\end{lstlisting}
\captionsetup{skip=2pt}
\caption{\rev{S2PL pseudocode for client and server sides.}}
\label{fig:2pl-pseudocode}
\end{figure}

\looseness=-1
To illustrate \ourframework's application, we analyze %start with 
a well-known distributed concurrency control protocol, namely Strict Two-Phase Locking (S2PL), which is commonly employed in settings with strong isolation and reliability requirements. 
In S2PL, each transaction acquires either exclusive write locks for writing to and reading from a key or shared read locks for just reading from a key. A transaction succeeds only if all locks for the involved keys are available. We assume that the failure to acquire any of these locks results in aborting the transaction \rev{(known as the "no-wait" method).} This protocol is usually combined with the Two-Phase Commit (2PC) protocol to achieve atomicity by ensuring that all servers involved in the transaction are prepared before committing.
\rev{\Cref{fig:2pl-pseudocode} shows the S2PL pseudocode from which we next construct a formal LTS protocol specification.} We then verify in \ourframework that it satisfies SSER.

\subsection{Protocol Specification}
\label{subsec:2pl-2pc-specification}

\rev{We first analyze the pseudocode's communication structure. The client broadcasts two (asynchronous) RPC requests to all servers involved in a transaction: (i) a prepare request, and (ii) a commit or abort request depending on the prepare responses. The servers send two RPC responses to the client: (i) an ``okay'' or ``not okay'' response to prepare, and (ii) a ``committed'' or ``aborted'' response.} 

\begin{figure}[t]
\begin{center}
   \includegraphics[width=.98\columnwidth]{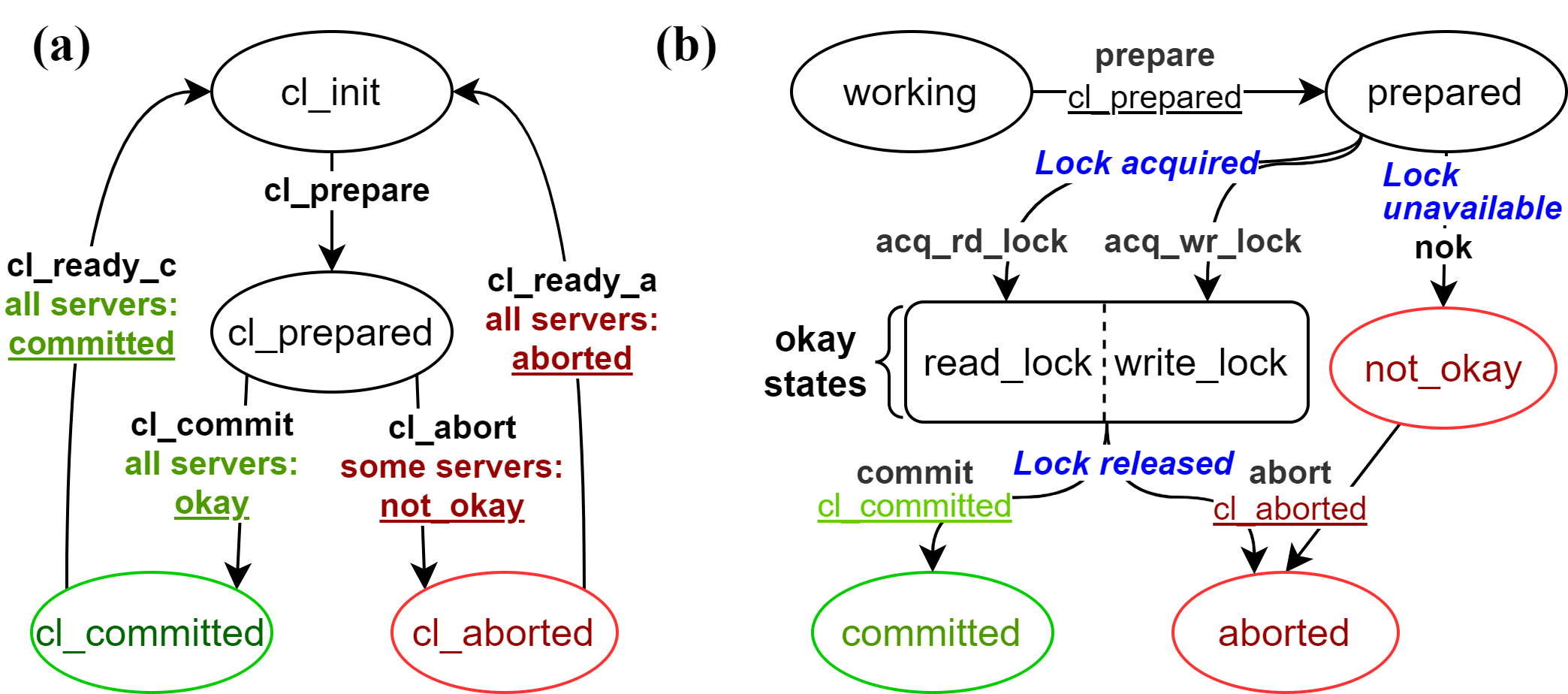}
\end{center}
\captionsetup{skip=5pt}
   \caption{S2PL: state diagrams of (a) a client's $\isaco{cl\_state}$ and (b) a server's $\isaco{svr\_state}$ for a given transaction. }
   \label{fig:2pl-diagrams}
   \vspace{-.7ex}
\end{figure}

\rev{We then translate this structure into control state diagrams of (a) a given client \inlisa{cl}'s %(coordinator) 
state and (b) a server's state for a given key \inlisa{k} and transaction \inlisa{t} (\Cref{fig:2pl-diagrams}).} The state transitions are labeled with event names (in bold) and the transition guards (underlined). 
\rev{The RPC requests and responses are each represented in the diagrams by an event on the sender and on the receiver side (\Cref{tab:RPC-to-event-mapping}). }

\rev{Recall from \Cref{subsec:protocol-modeling} that we model communication by the ``receiver'' reading the desired information from the configuration of the (passive) ``sender''.
Note that we can often combine receiving a message and sending the next message into one event. For example, the client commit event checks the servers' prepare responses and transitions to the \inlisa{cl\_committed} state (i.e., ``sends'' a commit message). Similarly, the server side commit and abort events handle both the RPC request and response. 
In contrast, the server prepare event only receives a prepare request and the response is provided by a lock-acquiring event or the \inlisa{nok} event.}
Next, based on the diagrams of \Cref{fig:2pl-diagrams}, we define the configurations and the events of our formal LTS protocol model of S2PL, called \tpl, in Isabelle/HOL. 
 
\begin{table}[t]
\captionsetup{skip=6pt}
\caption{\rev{Mapping RPC requests and responses to LTS events.}}
\small
\begin{tabular}{l|l|l}
    \toprule
     \textbf{RPC} & \textbf{client event} & \textbf{server event}\\\midrule
     \textbf{prepare request} & $\isaco{cl\_prepare}$ & $\isaco{prepare}$ \\\hline 
     \textbf{prepare response} & $\isaco{cl\_commit}$ or & $\isaco{acq\_rd\_lock}$ or  \\
     & $\isaco{cl\_abort}$ & $\isaco{acq\_wr\_lock}$ or $\isaco{nok}$  \\\hline
     \textbf{commit request} & $\isaco{cl\_commit}$ & $\isaco{commit}$\\\hline
     \textbf{commit response} & $\isaco{cl\_ready\_c}$ & $\isaco{commit}$ \\\hline
     \textbf{abort request} & $\isaco{cl\_abort}$ & $\isaco{abort}$\\\hline
     \textbf{abort response} & $\isaco{cl\_ready\_a}$ & $\isaco{abort}$  \\
     \bottomrule
\end{tabular}
\label{tab:RPC-to-event-mapping}
\end{table}

\subsubsection{Configurations}
\looseness=-2 We model the client, server, and global configurations as Isabelle/HOL records (\Cref{fig:2pl-configs}). 
The global configuration consists of mappings from client IDs to client configurations and from keys to server configurations. 
Each client maintains a transaction state (\inlisa{cl\_state}) and a sequence number (\inlisa{cl\_sn}).
%, and a view (\inlisa{cl\_view}). 
As clients are sequential, there is only one transaction state per client, with one of four values shown in \Cref{fig:2pl-diagrams}a. The sequence number keeps track of a client's current transaction. 
Each server responsible for a given key~\inlisa{k}, called server \inlisa{k} for short, stores the ongoing transactions' states  (\inlisa{svr\_state}), the key's version list (\inlisa{svr\_vl}), and the transactions' fingerprint on key~\inlisa{k} (\inlisa{svr\_fp}). 
Each transaction \inlisa{t} is in one of the seven states depicted in \Cref{fig:2pl-diagrams}b and its fingerprint \inlisa{svr\_fp(k,$\:$t)} on~\inlisa{k}, maps read and write operations to a value, if any.

%% NOTE: removed type variable 'v and replaced 'v in svr_fp by value
\begin{figure}[t]
\begin{lstlisting}
datatype txn_state =      -- transaction state
  cl_init | cl_prepared | cl_committed | cl_aborted

record cl_conf =          -- client configuration
  cl_state : txn_state
  cl_sn : nat

datatype ver_state =      -- version state
  working | prepared | read_lock | write_lock | 
  not_okay | committed | aborted

record svr_conf =         -- server configuration
  svr_state : txid $\fun$ ver_state
  svr_vl : list(version)
  svr_fp : txid $\times~\rwtype \map \valuetype$

record global_conf =      -- global configuration
  cls : cl_id $\fun$ cl_conf
  svrs : key $\fun$ svr_conf
\end{lstlisting}
\captionsetup{skip=3pt}
\caption{S2PL client, server, and global configurations.}
\label{fig:2pl-configs}
\end{figure}

\subsubsection{Events}
\looseness=-1 The client and server events change their respective transaction states as described in~\Cref{fig:2pl-diagrams}. 
%
%For brevity, 
We focus here on the server event \inlisa{acq\_write\_lock} and the client event \inlisa{cl\_commit}. The latter plays a central role in the refinement. Once a server \inlisa{k} is in the prepared state for some transaction \inlisa{t}, it tries to acquire a read or write lock for \inlisa{k}, depending on the intended operations. 
Here is the definition of the server event \inlisa{acq\_write\_lock}:

\begin{lstlisting}
definition acq_wr_lock(k, v$_w$, v$_r$, t) $\longleftrightarrow$
  -- guards:
  svr_state(k, t) = prepared $\land$
  $\forall$t'. not_locked(svr_state(k, t')) $\land$  $\;$ -- no locks on k
  v$_r$ $\in$ {$\bot$, last_ver_v(svr_vl(k))} $\land$ 
  -- updates:
  svr_state'(k) = write_lock $\land$ 
  svr_fp'(k, t) = [W $\mapsto$ v$_w$, R $\mapsto$ v$_r$]
\end{lstlisting}
%  -- remaining fields are unchanged
%  acq_wr_lock_unchanged s s'          -- rest unchanged
%
This event acquires a write lock for transaction \inlisa{t} to update the key~\inlisa{k} with the value \inlisa{v$_w$} and possibly also for reading the key's latest value \inlisa{v$_r$}, i.e., \inlisa{v$_r$} is either $\bot$ (no read) or \inlisa{last\_ver\_v$\;$(svr\_vl$\;$k)} (read) (Line~5). 
To execute the event, its guards require that 
(i) the server \inlisa{k} is in the \inlisa{prepared} state (Line~3), and 
(ii) no transaction has already acquired a lock on \inlisa{k} (Line~4). 
The event updates its state to \inlisa{write\_lock} (Line~7) and 
records its reads and writes in its key fingerprint variable \inlisa{svr\_fp} (Line~8). 

Once all servers have reached a locked state by executing a lock-acquiring event, the next event 
%to execute 
is the client's \inlisa{cl\_commit} event:

\begin{lstlisting}
definition cl_commit(cl, sn, u, F) $\longleftrightarrow$
  -- guards:
  cl_state(cl) = cl_prepared $\land$ 
  $\forall$k.$\:$dom(F(k))$\:\neq\:${}$\:\longrightarrow\:$is_locked(svr_state(k,$\:$get_txn(cl)))$\:\land$ 
  sn = cl_sn(cl) $\land$ 
  u = $\lambda$k. full_view($\refmapkvs{\_\tpl}$(s, k)) $\land$ 
  F = $\lambda$k. svr_fp(k, get_txn(cl)) $\land$ 
  -- updates:
  cl_state'(cl) = cl_committed  
\end{lstlisting}
% client view removed from client configs:
%  cl_view' cl = length $\circ$ (update_kv cl sn u F (kvs_of_gs s))
%  -- remaining fields are unchanged
%  cl_commit_unchanged s s'            -- rest unchanged

This event commits the client \inlisa{cl}'s current transaction with sequence number \inlisa{sn}, using the view \inlisa{u} and the transaction fingerprint~\inlisa{F}. The latter two parameters are only used for the refinement proof. 
The guards (Lines 3--7) require that 
(i) the client is in its prepared state, 
(ii) all involved servers are in a locked state for the client's transaction \inlisa{get\_txn(cl)}, 
(iii) \inlisa{sn} is the client's current sequence number,
(iv) \inlisa{u} corresponds to the \emph{full view} of the current abstract KVS  \inlisa{$\refmapkvs{\_\tpl}$(s)}, which includes all versions of all keys (see \Cref{subsec:2pl-2pc-refinement-proof}), and 
(v) \inlisa{F} is the complete fingerprint for the client's transaction constructed from all servers' per-key fingerprints, as set in the locking step. \rev{This includes the values of the keys read in the transaction.} 
The client's new state is \inlisa{cl\_committed} (Line~9). 

After the client's commit event, the transaction can no longer be aborted and all servers follow into their own commit state.

\subsection{Refinement Proof}
\label{subsec:2pl-2pc-refinement-proof}

To verify that the S2PL protocol $\tpl$ guarantees strict serializability, we prove in Isabelle/HOL that the concrete protocol model, as just specified, refines the abstract model $\isomodel{\SSER}$. 

\begin{theorem}%[Correctness of S2PL]
$\tpl \refines_{\refmapfull{r}{\pi}{\tpl}} \isomodel{\SSER}$.
\end{theorem}
\noindent In the remainder of this section, we sketch our proof of this theorem.

\subsubsection{Refinement mapping}
%\new{[shortened subsection to account for new subsection 3.3.1]}
We define the refinement mapping $r_{\tpl}$ on states, which consists of two components:
%, respectively reconstructing the abstract KVS and the client views: 
\begin{description}[leftmargin=10pt]
    \item[$\refmapkvs{\_\tpl}$:] 
    \looseness=-1 This function reconstructs the global KVS' version lists from all key servers' version lists, extended with the effects of ongoing transactions' \emph{client-committed} operations (cf.~\Cref{subsec:protocol-verification-technique}). 
    
    \item[$\refmapviews{\_\tpl}$:] 
    Since the client commit event is always using the \emph{full view}, there is no need to store a concrete view in the client configurations. This function thus maps all concrete configurations to the (dummy) initial view $\lambda k.\, \{0\}$ of the abstract model. 
\end{description}

%\subsubsection{Mediator Function}
The refinement mapping $\refmap{\pi}{\tpl}$ on events maps \tpl{}'s \inlisa{cl\_commit} event to the abstract commit event and all other events to the abstract $\Skip$ event.
%, which leaves the abstract configuration unchanged. 
To reconstruct the abstract commit event's parameters from the \inlisa{cl\_commit} event's parameters, we have added the former, namely, the intermediate view~\inlisa{u}, the fingerprint~\inlisa{F}, and the sequence number~\inlisa{sn} to \inlisa{cl\_commit}'s event parameters.
Note that no event of \tpl{} refines the abstract view extension event.

\subsubsection{Proof of refinement (sketch)}
%Besides the condition for the initial state, 
The main part of the proof consists of showing that for any concrete transition $s \myrightarrow{e} s'$ 
%from configuration $s$ to $s'$ labeled with an event $e$, 
there is a transition $\refmap{r}{\tpl}(s)$ \raisebox{-1.5pt}{$\myrightarrow{\refmap{\pi}{\tpl}(e)}$} $\refmap{r}{\tpl}(s')$ in 
%the abstract model 
$\isomodel{\SSER}$. 
For all events other than \inlisa{cl\_commit}, we must show that $\refmap{r}{\tpl}(s) = \refmap{r}{\tpl}(s')$, which corresponds to an abstract $\Skip$. This is the easy part.
For the event \inlisa{cl\_commit(sn,$\;$u,$\;$F)}, we divide the proof that it refines the abstract commit event with the same parameters (cf.~\Cref{subsec:abstract-model}) into guard strengthening, i.e., that the concrete event's guards imply the abstract guards, and update correspondence (cf.~\Cref{subsec:lts-refinement}).
Let $\kvs = \refmapkvs{\_\tpl}(s)$ and $\kvs' = \refmapkvs{\_\tpl}(s')$. 

\looseness=-1 For guard strengthening, we focus on two guards here. 
(i) The guard $\canCommit(\kvs, \inlisa{u}, \inlisa{F}, R_{\SSER})$ requires that the view \inlisa{u} is \emph{closed} under the relation $R_{\SSER} = \WW_{\kvs}^{-1}$, which means that all versions must be visible. \ourframework provides a general, reusable lemma showing that this is the case for full views.
(ii) The guard $\LWW(\kvs, \inlisa{u}, \inlisa{F})$ requires that the values read by the transactions as indicated by the fingerprint \inlisa{F} equal the latest values visible in the (full) view~\inlisa{u} of~$\kvs$. This is the most interesting case. 
Our S2PL protocol model \tpl{} satisfies this condition since the strict locking discipline ensures that, for each key, at most one transaction at a time can write a new version to the KVS. In particular, there cannot be any concurrent client-committed, but not yet server-committed writes to a key that the client's transaction reads. Proving this guard requires several invariants about locks and their relation to the fingerprint~\inlisa{F}, discussed below.

For update correspondence, we must establish that $\kvs'$ is equal to \inlisa{UpdateKV($\kvs$,Tn(cl,sn),u,F)}, which requires a series of lemmas about state updates. These in turn rely on \ourframework's library lemmas about abstract state updates.

\subsubsection{Required invariants and lemmas.} 
%\new{[shortened subsection to account for new subsection 3.3.1]}
The refinement proof uses numerous invariants and lemmas that describe the system's behaviour. 
First, it uses the generic invariants from \Cref{subsubsec:protocol-invariants}, except view closedness, which trivially holds for the full views used in S2PL. 
Additionally, we need the following locking-related invariants for S2PL. These can be adapted to other lock-based protocols.
\begin{description}[leftmargin=10pt]
    
    \item[Lock invariants] These invariants capture the multiple-reader, single-writer semantics of read and write locks. 
    
    \item[Fingerprint invariants] \looseness=-1 Three invariants relate each locking scenario (read or write) to the expected fingerprints on key servers.     
    
\end{description}

\section{Finding Isolation Bugs using \ourframework}
\label{sec:tapir}

In this section, we demonstrate how \ourframework can also help uncover isolation bugs in protocols. 
\rev{We begin with an overview of our approach and then present a case study on the TAPIR protocol~\cite{DBLP:journals/tocs/ZhangSSKP18}, 
where we uncover a previously unknown atomic visibility violation.}

\subsection{\rev{Approach}}
\label{sec:isobug-approach}

\looseness=-1
\rev{We discover isolation bugs from failed protocol verification attempts. Hence, we start by %modeling the protocol and 
setting up a refinement proof for the claimed isolation guarantee in \ourframework (cf.~\Cref{subsec:protocol-verification-technique}).
If we are unable to prove a given proof obligation or required invariant, even after potential improvements to the model and the refinement mappings, this may indicate a problem with the protocol. We then analyze the situation further and try to construct a counterexample protocol execution that violates the desired isolation guarantee. }

\rev{In the following, we consider different cases in such an analysis, focusing on the most interesting case, namely, the failure to establish guard strengthening for events that were intended to refine the abstract commit event. 
We start from the basic abstract guards, shared by all isolation levels, and progress to the more complex ones, specific to a given isolation level. For each category of guards, we assume the preceding categories' guards to be already proven.}
Our step-by-step approach of breaking down such proofs into small and manageable proof obligations, corresponding to different protocol aspects, facilitates spotting isolation bugs in protocol designs.

\subsubsection{\rev{For RA: View wellformedness and view extension}}
\rev{If a wellformedness guard fails because of view atomicity, either the protocol inherently does not support atomic views, which immediately shows a violation of RA, or the refinement mapping must be updated to include all versions written by a given transaction in the view. Otherwise, if the guard fails due to indices pointing to non-existent versions, again the refinement mapping's view construction must be checked. The view extension guard's failure indicates that some versions, previously visible to the client, are removed from its view, which can be directly checked in the protocol definition or again indicates a need for updating the refinement mapping.}

\subsubsection{\rev{For RA: Last-write-wins $(\LWW)$}}  

\rev{When $\LWW$ fails, a client might not read the latest version in its view. An important special case of this occurs when the client reads a transaction's update on one server but reads an older version on another server that was also updated by that transaction, despite having both updates in its view (as view atomicity is assumed proven). This behavior corresponds to the fractured read anomaly, which indicates the violation of RA.} 

\begin{figure}[t]
\hspace{-5pt}
\includegraphics[width=1\columnwidth]{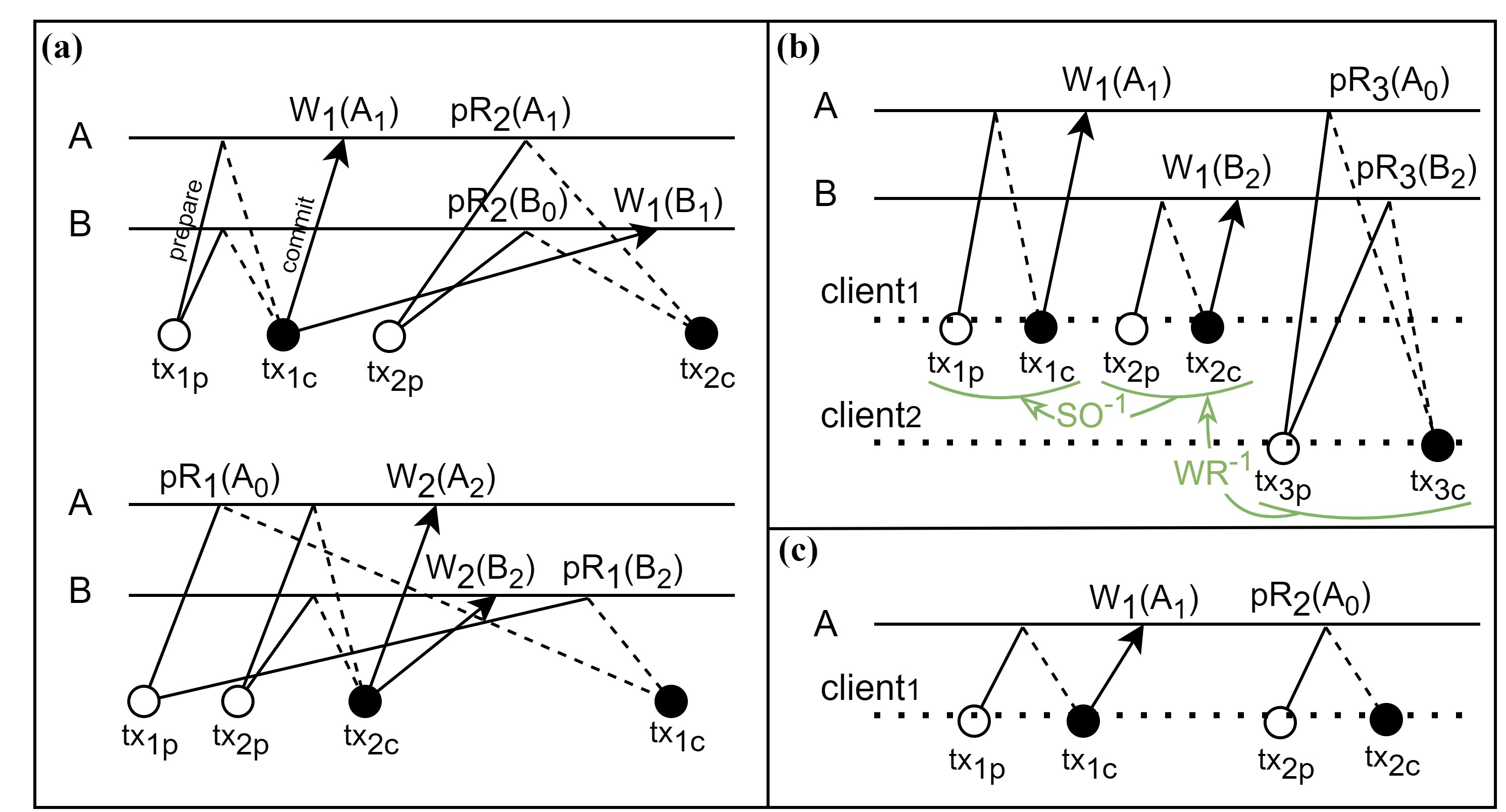}
   \caption{\rev{Interleaving of transactions' events, violating (a)~$\LWW$, (b)~$\canCommit_{\TCC}$, and (c)~$\vShift_{\isaco{RYW}}$. Clients' \emph{prepare} and \emph{commit} events are shown by the $\mathtt{tx_{\_p}}$ and $\mathtt{tx_{\_c}}$ nodes. 
   Servers' prepare-read and commit-write events are marked by $\mathtt{pR_{txid}(read{\text -}version)}$ and $\mathtt{W_{txid}(written{\text -}version)}$ on keys.}}
   \label{fig:tapir_cntr_ex_PO}
\end{figure}

\rev{To construct an execution that captures this failure, at least two transactions are required: A write transaction, writing to at least two keys, and a read transaction that does not read the latest version on at least one of those keys. 
\Cref{fig:tapir_cntr_ex_PO}a shows two possible interleavings of such transactions, both violating the $\LWW$ condition and exhibiting fractured reads.} 

\subsubsection{\rev{For higher isolation levels: $\canCommit$ and $\vShift_{\isolevel}$}}
\looseness=-1 \rev{If all previous proof obligations are proven, a failure in these guards can indicate an isolation bug for the given $\isolevel$. In these cases, similar to the $\LWW$ case, we must construct executions that capture these failures. For $\canCommit$, by starting from an interleaving of transactions, and considering a client's view $u$, which defines the set of transactions visible to it ($\visTx(\kvs,u)$), we must repeatedly apply $R_{\isolevel}^{-1}$ to this set until the result contains an \emph{invisible} transaction, i.e., a transaction that is neither read-only nor in $u$. An example of such an execution for $\isolevel = \TCC$ is shown in \Cref{fig:tapir_cntr_ex_PO}b. For $\vShift$, the session guarantees (monotonic reads or read-your-writes) fail. Therefore, the execution should demonstrate the failure of these guarantees. \Cref{fig:tapir_cntr_ex_PO}c shows an example execution for violating read-your-writes.}

\smallskip
\looseness=-1 \rev{After building a candidate execution that violates a proof obligation, we might need to assign some values, e.g., timestamps, or add transactions, such that the protocol can run the given execution. We use the protocol's model to infer these values and needed additions. 
We then test the final execution against the protocol specification to ensure it is a valid protocol execution.
Finally, we check if the resulting counterexample is indeed a violation of an isolation guarantee.}

\subsection{Case Study: TAPIR Protocol}
\label{sec:isobug-tapir}

\begin{figure}[t]
\begin{lstlisting}
datatype txn_state =                -- transaction state
  cl_prepared (ts, key $\map$ txid, key $\map \valuetype$) |
  cl_committed (ts, key $\map$ txid, key $\map \valuetype$) |
  ... -- other constructors: same as S2PL

record cl_conf =                    -- client configuration
  cl_state : txn_state
  cl_sn : nat
  cl_local_time : ts

datatype ver_state =                -- version state
  prepared (ts, txid$_\bot$, $\valuetype_\bot$) |
  committed (ts, txid$_\bot$, $\valuetype_\bot$) |
  ... -- other constructors: same as S2PL

record svr_conf =                   -- server configuration
  svr_state : txid $\fun$ ver_state

record global_conf =                -- global configuration
  cls : cl_id $\fun$ cl_conf
  svrs : key $\fun$ svr_conf
  commit_order : key $\fun$ list(txid)  -- a history variable
\end{lstlisting}
\captionsetup{skip=3pt}
\caption{\rev{TAPIR client, server, and global configurations.}}
\label{fig:tapir-configs}
\end{figure}

\looseness=-1
TAPIR is a distributed transaction protocol, which combines 2PC with optimistic concurrency control (OCC) and claims to offer \emph{strictly serializable} (SSER) read-write transactions.
Before committing a transaction, TAPIR  performs a \emph{timestamp-based} OCC validation, checking for any concurrent conflicting transactions. 
%It is claimed to provide \emph{strictly serializable} read-write transactions.
Moreover, by building it on an \emph{inconsistent replication} protocol, the entire transaction system offers fault tolerance while eliminating the coordination redundancy found in conventional replicated databases 
%(e.g., ~\cite{corbett2013spanner}).
like Spanner~\cite{corbett2013spanner}.

%{Modeling TAPIR in \ourframework} 
\looseness=-1 We model TAPIR similar to S2PL, except that (i) TAPIR servers perform OCC checks instead of locking to prepare a read or write, 
%\rev{(ii) read and write operation results are stored directly in the version states instead of maintaining a version list,} 
\rev{(ii) operation results are stored directly in the version states instead of keeping version lists or key fingerprints,} 
and (iii) the client's state contains a local clock for generating timestamps. 
\rev{These local clocks are \emph{loosely synchronized}, and the authors state that the protocol only depends on their synchronization for performance but not for correctness. \Cref{fig:tapir-configs} shows the TAPIR protocol's configurations.}

\begin{figure}[t]
\begin{lstlisting}
definition prepared_rd_tstmps(k) $\longleftrightarrow$
  {ts$\:|\:$svr_state(k,$\;$t)$\;=\;$prepared(ts,$\;$r,$\;$w)$\;\land\;$r$\;\neq\bot$}

definition prepared_wr_tstmps(k) $\longleftrightarrow$
  {ts$\;|\;$svr_state(k,$\;$t)$\;=\;$prepared(ts,$\;$r,$\;$w)$\;\land\;$w$\;\neq\bot$}

definition committed_wr_tstmps(k) $\longleftrightarrow$
  {ts$\;|\;$svr_state(k,$\;$t)$\;=\;$committed(ts,$\;$r,$\;$w)$\,\land\:$w$\;\neq\bot$}

definition tapir_occ_check(k, t, ts, t$_r$, v$_w$) $\longleftrightarrow$
  if t$_r$ $\neq \bot$ $\land$ committed_wr_tstmps(k) $\neq$ {} $\land$
     ver_ts(svr_state(k,$\;$t$_r$)) < Max(committed_wr_tstmps(k))
  then aborted
  else if t$_r$ $\neq \bot$ $\land$ prepared_wr_tstmps(k) $\neq$ {} $\land$              
          ts > Min(prepared_wr_tstmps(k))
  then aborted     -- originally: ABSTAIN
  else if v$_w$ $\neq \bot$ $\land$ prepared_rd_tstmps(k) $\neq$ {} $\land$             
          ts < Max(prepared_rd_tstmps(k))
  then aborted     -- originally: RETRY
  else if v$_w$ $\neq \bot$ $\land$ committed_wr_tstmps(k) $\neq$ {} $\land$
          ts < Max(committed_wr_tstmps(k))
  then aborted     -- originally: RETRY
  else prepared(ts,$\;$t$_r$,$\;$v$_w$)
\end{lstlisting}
\captionsetup{skip=0pt}
\caption{\rev{TAPIR OCC check, modeled after \cite{DBLP:journals/tocs/ZhangSSKP18}'s definition.}}
\label{fig:tapir-occ-check}
\vspace{-.7ex}
\end{figure}

\begin{figure}[t]
\begin{lstlisting}[firstnumber=14]
  else if t$_r$ $\neq \bot$ $\land$ prepared_wr_tstmps(k) $\neq$ {} $\land$
    ver_ts(svr_state(k,$\;$t$_r$)) < Min(prepared_wr_tstmps(k))
\end{lstlisting}
\captionsetup{skip=0pt}
\caption{\rev{TAPIR OCC check's different condition for preparing reads, based on the (older) conference definition~\cite{tapir}.}}
\label{fig:tapir-occ-check-conf}
\end{figure}

\looseness=-1 
\rev{\Cref{fig:tapir-occ-check} presents our modeling of the \isaco{tapir\_occ\_check} function from~\cite{DBLP:journals/tocs/ZhangSSKP18} in \ourframework. 
The function's arguments are the key~\isaco{k}, transaction ID~\isaco{t}, the proposed timestamp~\isaco{ts}, the read version's transaction ID~\isaco{t$_r$}, and the written value~\isaco{v$_w$} ($\bot$ for t$_r$ or v$_w$ indicates no reads or writes on key~\isaco{k}). The check results in a new server state, either \inlisa{aborted} or \inlisa{prepared}. 
Note that TAPIR's OCC check returns ABSTAIN or RETRY in certain cases (see comments in \Cref{fig:tapir-occ-check}). These are used to improve performance by trying to avoid certain unnecessary aborts that would otherwise arise due to TAPIR's interaction with the underlying (inconsistent) replication protocol. As we do not model replication, we map both of these to \inlisa{aborted}, which is safe in that it cannot introduce any new violations of SSER.}

\rev{There is a different, earlier version of the OCC check~\cite{tapir}, which corresponds to replacing one of the conditions for preparing new reads (Lines~14 and~15 in \Cref{fig:tapir-occ-check}) by the two lines in \Cref{fig:tapir-occ-check-conf}. 
%The implementation [REF] uses the version from ~\cite{DBLP:journals/tocs/ZhangSSKP18}, modeled in \Cref{fig:tapir-occ-check}.
TAPIR's codebase~\cite{tapir-code} uses the newer definition in~\cite{DBLP:journals/tocs/ZhangSSKP18}~(\Cref{fig:tapir-occ-check}).}

\subsection{Finding TAPIR Isolation Bugs using \ourframework}
\label{sec:isobug-find}

\looseness=-1 Lu et al.~\cite{osdi23} have found an issue with SSER's real-time ordering in TAPIR but conjecture that it still satisfies SER.
Using \ourframework, we discover that TAPIR violates \emph{atomic visibility} (RA), a much weaker isolation level than (S)SER.
\rev{This bug escaped both the TAPIR authors' manual correctness proofs (including SSER) and their additional TLA+ model checking analysis.}
This underscores the need for our \emph{rigorous} verification approach, which covers \emph{all} system behaviors. 

\looseness=-1 
\rev{As TAPIR aims to provide $\SSER$,} we set up a corresponding refinement proof in \ourframework. \rev{The mapping $\pi$ maps the client commit event to the abstract commit, and all others to $\Skip$. The state mapping $r$ reconstructs the abstract KVS's version lists by mapping the transaction IDs in $\isaco{commit\_order}$ (recorded in the order of client commits) to the corresponding abstract versions extracted from their version state. As for S2PL, the abstract view is the constant initial view.}

For $\SSER$, \rev{the view \inlisa{u} used in the commit event must contain all abstract versions. }
%For the $\SSER$ isolation level, the view must stay closed under $R_{\SSER} = \WW_{\kvs}^{-1}$, i.e., each client's view contains all client-committed versions. 
Using general lemmas about full views, we prove the abstract guards for view wellformedness~($\wellformed$), \rev{extension~($\views(cl) \sqsubseteq u$),} and closure ($\canCommit$). 
However, we could not prove the abstract $\LWW$ guard, \rev{despite following the steps in \Cref{subsubsec:refinement-proofs}. 
As $\LWW$ must hold for any isolation level of our abstract model, this indicates that TAPIR potentially violates atomic visibility (RA). 
We confirm this by exhibiting two witnessing counterexamples, one for each version of the OCC check.} These remain valid when TAPIR is layered on top of the replication protocol.

\begin{figure}[t]
\begin{center}
   \hspace{-8pt}\includegraphics[width=1.03\columnwidth]{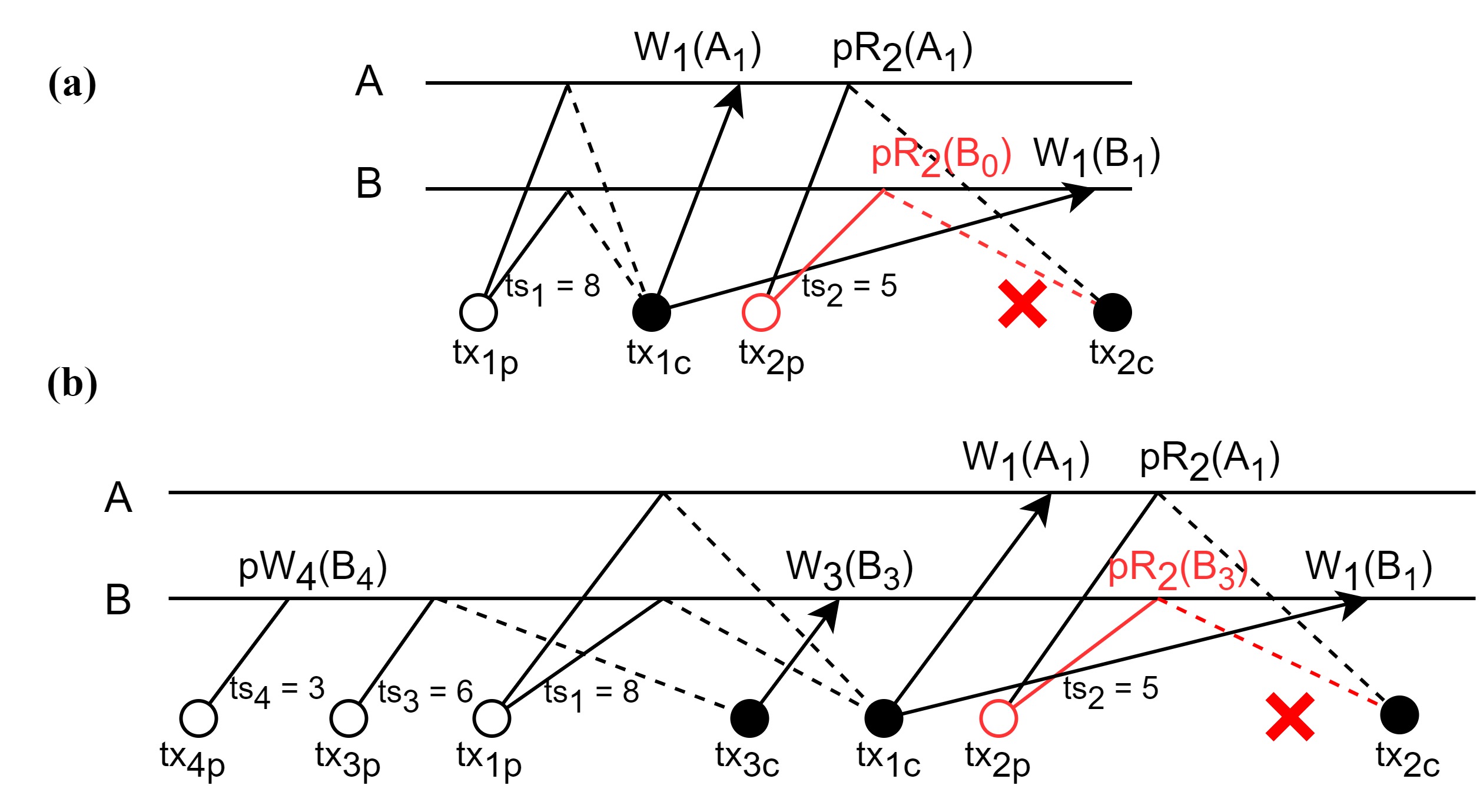}
\end{center}
\captionsetup{skip=1pt}
   \caption{\looseness=-1 Atomic visibility counterexamples for TAPIR, based on $\inlisa{tapir\_occ\_check}$'s definition in (a)~\cite{DBLP:journals/tocs/ZhangSSKP18} and (b)~\cite{tapir}.
   }
   \label{fig:tapir_cntr_ex}
   \vspace{-.3ex}
\end{figure}

\begin{example}[journal version~\cite{DBLP:journals/tocs/ZhangSSKP18}]
\looseness=-1 \rev{We start from the transactions' interleaving shown in \Cref{fig:tapir_cntr_ex_PO}a (top) and try assigning timestamps to each transaction such that they can prepare their reads and writes. 
The write transaction $tx_1$ can prepare its writes with any proposed timestamp $ts_1>0$ because initially there are no committed writes or prepared reads. Hence, the $\isaco{tapir\_occ\_check}(k,tx_1,ts_1,\bot,v)$ calls for $k\in\{A,B\}$ return $\isaco{prepared}$, as they are not aborted by any of the checks in \Cref{fig:tapir-occ-check}. The read transaction $tx_2$ calls $\isaco{tapir\_occ\_check}(k,tx_2,ts_2,t_r,\bot)$, for $k\in\{A,B\}$. Since at the execution time of these server prepare events only $A_1$ is committed, the maximum of committed writes' timestamps are $ts_1$ and $0$ on $A$ and $B$ respectively, which are equal to the timestamps of the versions read by $t_2$, i.e., they are not aborted at Line 13. Furthermore, $B$ has one prepared write ($B_1$), so $ts_2\leq ts_1$ must hold to prevent $pR_2(B_0)$ from aborting on Line 16.}
Therefore, as shown in \Cref{fig:tapir_cntr_ex}a, by assigning $ts_1 = 8$ and $ts_2 = 5$, \rev{we get a valid execution of TAPIR} that violates atomic visibility, as $tx_1$'s writes are only partially observed by $tx_2$.
\end{example}

\begin{example}[\rev{conference version~\cite{tapir}}]
\looseness=-1 
\rev{We modify our previous counterexample to obtain a counterexample for the older definition (\Cref{fig:tapir-occ-check-conf}), where the read version's timestamp (denoted by $\isaco{ver\_ts(svr\_state}(k,t_r))$) instead of the transaction's proposed timestamp is checked for preparing reads.
In \Cref{fig:tapir_cntr_ex}a, at the time of $tx_2$'s prepare read events, there is one prepared write on $B$ with $ts_1 = 8$, which means the prepare read on $B$ would abort on Line~16 ($0<Min\{8\}$). So the read transaction should read a slightly newer version on $B$ (see \Cref{fig:tapir_cntr_ex}b), for example, a single write by $tx_3$ on $B$ with $ts_3 = 6$ (still older than $W_1(B_1)$), and also have another transaction $tx_4$ prepare a write on $B$ with a lower timestamp, e.g., $ts_4 = 3$, so that it pulls the minimum down and $6\nless Min\{3, 8\}$ prevents 
%the event from
aborting on Line~16. We then check if the two new write transactions also pass the checks. As these transactions' prepare events appear before any prepare\_read or commit\_write events, the corresponding sets are empty, so both transactions successfully prepare their writes. \Cref{fig:tapir_cntr_ex}b shows the completed example, depicting a valid execution of TAPIR that violates atomic visibility, as $tx_1$'s writes are only partially observed by $tx_2$.}

\end{example}

\subsubsection*{\rev{Validation}}
\rev{\looseness=-1 We confirmed the violation of atomic visibility in TAPIR's implementation~\cite{tapir-code}. 
We treated TAPIR as a black box to preserve its codebase and minimize the risk of introducing errors, and collected transaction execution histories from its runs %For these runs, we employed TAPIR's own workload generator with a small configuration with 3 clients, 2 shards, 2 keys, 2 operations per transaction, and 50\% reads. (is already in appendix)
%We then analyzed the collected histories 
and analyzed them using IsoVista~\cite{isovista}, a tool for detecting isolation anomalies across various isolation levels. % including RA. 
Its RA-checking component identified violations of atomic visibility in these histories (see \iffull \Cref{app:validation} \else \cite[Appx. A]{tech-rpt}\fi).
}

\section{Related Work}

Along with the advances in designing reliable, performant database transactions, considerable efforts have been dedicated to formalizing their isolation guarantees. We classify them into two categories.

\inlsec{Declarative Specifications}
\emph{Dependency graphs} were 
%first 
introduced by Bernstein et al.~\cite{bernstein1981concurrency} for characterizing serializability and
%in early 1980s. They were 
later adapted by Adya to weaker isolation levels such as SI and RC~\cite{adya1999weak}.
%snapshot isolation and read committed by Adya.

Based on \emph{abstract executions}, Cerone et al.~\cite{DBLP:conf/concur/Cerone0G15} propose
an axiomatic framework to declaratively define
 isolation levels at least as strong as read atomicity,
with the dual notions of visibility (what
transactions can observe) and arbitration (the order of installed
versions/values).
To ease the validations of isolation guarantees over database execution histories,
Biswas and Enea~\cite{DBLP:journals/pacmpl/BiswasE19} present an alternative axiomatic framework for characterizing isolation criteria via the write-read relation which,  in contrast to the visibility
relation~\cite{DBLP:conf/concur/Cerone0G15}, can be extracted straightforwardly from histories.

\looseness=-1
More recently,
%Recent years have seen the development of 
effective isolation checkers have been developed based on such % some of these 
declarative specifications~\cite{DBLP:journals/pacmpl/BiswasE19,elle,cobra,polysi,isovista,plume}.
These include Biswas and Enea's 
dbcop~\cite{DBLP:journals/pacmpl/BiswasE19}, 
%based on their axiomatic framework
and 
Elle~\cite{elle}, based on dependency graphs~\cite{adya1999weak}, 
for checking multiple isolation levels,
and
Cobra~\cite{cobra} for SER
and PolySI~\cite{polysi} for SI. 
All these tools verify whether individual database execution histories satisfy the required isolation level. 
Hence, they
can only show the presence of errors, not their absence, in database implementations.
Given that many isolation bugs stem
from design-level defects in carefully designed and heavily tested production databases, full verification of database designs is highly desirable. Our \ourframework framework fulfills this requirement.

\inlsec{Operational Specifications}
\looseness=-1 
Crooks et al.~\cite{seeing} introduce a state-based formalization of isolation levels based on complete transaction traces; 
\ourframework only uses the current database state and clients' views on that state.
Liu et al.~\cite{DBLP:conf/tacas/LiuOZWM19} develop a formal framework in %the 
Maude~\cite{AllAboutMaude:Book2007} 
%language 
for 
specifying and model checking isolation properties of a database execution history.
%of transactions. 
%collected from executing a transaction system.
Lesani et al.~\cite{LesaniBC16} propose an operational framework for the verification of causally-consistent KVSs. 

We base our framework \ourframework on the operational semantics of isolation guarantees in~\cite{DBLP:conf/ecoop/XiongCRG19}. 
In contrast to the aforementioned related work, \ourframework supports \emph{invariant-based verification}, which can be mechanized using theorem provers like Isabelle/HOL. 
Most related works~\cite{bernstein1981concurrency,adya1999weak,DBLP:conf/concur/Cerone0G15,seeing,DBLP:conf/ecoop/XiongCRG19} only provide theories of isolation criteria with no tool support for formal analysis. Similar to \ourframework, the framework in~\cite{LesaniBC16} is formalized in in a theorem prover, namely, Coq. However, it only supports non-transactional KVSs and causal consistency. 
The framework in~\cite{DBLP:conf/tacas/LiuOZWM19} supports transactional databases and is accompanied by an explicit-state model checker. 
However, it offers weaker correctness guarantees, limited to a small number of processes and transactions.
In contrast, \ourframework supports rigorous verification of arbitrary numbers of processes and transactions.

\section{Discussion and Conclusion}

%\rev{We take a step back to review and discuss our approach. }
%
%\Cref{tab:isabelle-stats} shows some statistics of our formalization effort. 

\inlsec{Framework Infrastructure and Guarantees} 
\ourframework covers a wide spectrum of isolation levels ranging from RA to SSER and can thus be applied to a large variety of protocols. 
\ourframework provides both methodological guidance and structural support for protocol %modeling and
verification, \rev{including a protocol modeling style and a verification methodology and infrastructure. The latter consists of a sizable collection of supporting lemmas and proof rules (cf.~first row of \Cref{tab:isabelle-stats}).} 
Our correctness proofs provide strong guarantees, as they cover arbitrary numbers of processes and transactions, while model checking and testing can only \rev{find bugs, but not prove their absence.} 

\begin{table}[t]
\footnotesize
    \centering
    \captionsetup{skip=6pt}
    \caption{Statistics of the Isabelle/HOL formalization. 
    }
    \label{tab:isabelle-stats}
    \begin{tabular}{l r r r r r}
    \toprule
                & \textbf{Tot LoC} & \textbf{Model LoC} & \textbf{Proof LoC} & \textbf{\#lemmas} & \textbf{\#inv.}  \\ \midrule    
     \ourframework framew.    &  3093  &  --     &   --  &   275     &     11       \\    
     S2PL+2PC                 &  1890  &   285   & 1605  &   70     & 13 \\
     TAPIR                    &   951  &   386   &  565  & 14  &  12 \\
     \midrule 
     \textbf{Total} & 5934 & 671 & 2170 & 359 & 36\\
     \bottomrule
     
    \end{tabular}
%    \vspace{-1ex}
\end{table}

\inlsec{\rev{Modeling and Verification Effort}} 
\rev{Protocol modeling requires relatively little effort (e.g., a few days) and can be done by developers with some experience in functional programming. 
However, protocol verification using an interactive prover like Isabelle requires considerable expertise. The effort depends on the protocol and may range from a few weeks to several months. }

\inlsec{\rev{Reusability}} 
\looseness=-1
\rev{The abstract model can easily be instantiated to additional isolation levels stronger than RA, given a suitable dependency relation $R_{IL}$ and session guarantee $\vShift_{IL}$. 
For protocol modeling, we can reuse the top-level structure of configurations and the 2PC structure of events.
For verification,
we have identified several %useful 
categories of invariants reusable or adaptable to other protocol verification efforts. 
Protocols designed for particular isolation levels come with their own types of invariants. For example, in~\cite{GhasemiradSprengerLiu+-TACAS25}, we have verified a TCC protocol, which required numerous invariants about timestamps. We expect these to be reusable for other TCC protocols.}

\inlsec{\rev{Limitations}}
\looseness=-1
\rev{Our framework is based on the KVS model. It supports neither general relational databases nor predicate queries. Doing so would  require a study of the formal semantics of weak isolation levels in such a setting. % and substantial changes to our model.
We also do not model any kind of failures. In particular, the 2PC protocol underlying most protocols is itself blocking. Some fault tolerance might be achievable using replication, which we consider to be orthogonal, as mentioned earlier.}

\inlsec{Conclusions and Future Work}
\ourframework is the first systematic, mechanized framework for formally verifying that database transaction protocols \rev{conform to their intended isolation levels.} We demonstrated its effectiveness through \rev{two case studies, involving both verification and falsification.}
Given repeated occurrences of isolation bugs in database protocol designs, our work can help design more reliable transaction systems.

%\inlsec{Future Work}
\looseness=-1
We plan to extend \ourframework with additional isolation guarantees, including weaker levels such as RC.
We also intend to facilitate protocol invariant proofs, for example, by using techniques for automated invariant generation, and refinement proofs, by introducing a generic distributed protocol model between the current centralized abstract model and concrete protocol models. This intermediate model could factor out recurring parts of protocol refinement proofs.
Moreover, extending the correctness guarantees from designs to implementations would be desirable, e.g., using the approach in~\cite{SprengerKEWMCB2020}.

%% previous version of future work
%%
%We plan to extend \ourframework with \rev{additional isolation guarantees, including weaker levels such as RC, and with additional infrastructure to facilitate refinement proofs. 
%
%For example, we could refine our abstract transaction model into a generic abstract protocol model capturing the 2PC structure of many protocols, thus factoring out recurring parts of refinement proofs. This should facilitate the proof that concrete protocol models refine the abstract protocol model.}
%Moreover, extending the correctness guarantees from designs to implementations would be desirable, e.g., using the approach in~\cite{SprengerKEWMCB2020}.

\begin{acks}
    We thank the anonymous reviewers for their valuable feedback. This research is supported by an ETH Zurich Career Seed Award and the Swiss National Science Foundation project 200021-231862 ``Formal Verification of Isolation Guarantees in Database Systems''.
\end{acks}

%\clearpage
\bibliographystyle{ACM-Reference-Format}
\bibliography{VerIso}


\begin{thebibliography}{58}

%%% the Web address.

\ifx \showCODEN    \undefined \def \showCODEN     #1{\unskip}     \fi
\ifx \showDOI      \undefined \def \showDOI       #1{#1}\fi
\ifx \showISBNx    \undefined \def \showISBNx     #1{\unskip}     \fi
\ifx \showISBNxiii \undefined \def \showISBNxiii  #1{\unskip}     \fi
\ifx \showISSN     \undefined \def \showISSN      #1{\unskip}     \fi
\ifx \showLCCN     \undefined \def \showLCCN      #1{\unskip}     \fi
\ifx \shownote     \undefined \def \shownote      #1{#1}          \fi
\ifx \showarticletitle \undefined \def \showarticletitle #1{#1}   \fi
\ifx \showURL      \undefined \def \showURL       {\relax}        \fi
% invisible to TeX
\providecommand\bibfield[2]{#2}
\providecommand\bibinfo[2]{#2}
\providecommand\natexlab[1]{#1}
\providecommand\showeprint[2][]{arXiv:#2}

\bibitem[Abadi and Lamport(1991)]%
        {AL91}
\bibfield{author}{\bibinfo{person}{Mart{\'{\i}}n Abadi} {and} \bibinfo{person}{Leslie Lamport}.} \bibinfo{year}{1991}\natexlab{}.
\newblock \showarticletitle{The Existence of Refinement Mappings}.
\newblock \bibinfo{journal}{\emph{Theor. Comput. Sci.}} \bibinfo{volume}{82}, \bibinfo{number}{2} (\bibinfo{year}{1991}), \bibinfo{pages}{253--284}.
\newblock
\urldef\tempurl%
\url{https://doi.org/10.1016/0304-3975(91)90224-P}
\showDOI{\tempurl}


\bibitem[Adya(1999)]%
        {adya1999weak}
\bibfield{author}{\bibinfo{person}{Atul Adya}.} \bibinfo{year}{1999}\natexlab{}.
\newblock \emph{\bibinfo{title}{Weak consistency: a generalized theory and optimistic implementations for distributed transactions}}.
\newblock \bibinfo{thesistype}{Ph.\,D. Dissertation}. \bibinfo{school}{Massachusetts Institute of Technology, Department of Electrical Engineering and Computer Science}.
\newblock


\bibitem[Akkoorath et~al\mbox{.}(2016)]%
        {Cure:ICDCS2016}
\bibfield{author}{\bibinfo{person}{Deepthi~Devaki Akkoorath}, \bibinfo{person}{Alejandro~Z. Tomsic}, \bibinfo{person}{Manuel Bravo}, \bibinfo{person}{Zhongmiao Li}, \bibinfo{person}{Tyler Crain}, \bibinfo{person}{Annette Bieniusa}, \bibinfo{person}{Nuno~M. Pregui{\c{c}}a}, {and} \bibinfo{person}{Marc Shapiro}.} \bibinfo{year}{2016}\natexlab{}.
\newblock \showarticletitle{Cure: Strong Semantics Meets High Availability and Low Latency}. In \bibinfo{booktitle}{\emph{36th {IEEE} International Conference on Distributed Computing Systems, {ICDCS} 2016}}. \bibinfo{publisher}{{IEEE} Computer Society}, \bibinfo{pages}{405--414}.
\newblock
\urldef\tempurl%
\url{https://doi.org/10.1109/ICDCS.2016.98}
\showDOI{\tempurl}


\bibitem[Azure(2022)]%
        {Azure}
\bibfield{author}{\bibinfo{person}{Azure}.} \bibinfo{year}{2022}\natexlab{}.
\newblock \bibinfo{title}{azure-cosmos-tla}.
\newblock
\newblock
\urldef\tempurl%
\url{https://github.com/Azure/azure-cosmos-tla}
\showURL{%
Retrieved March~7, 2025 from \tempurl}


\bibitem[Bailis et~al\mbox{.}(2016)]%
        {ramp}
\bibfield{author}{\bibinfo{person}{Peter Bailis}, \bibinfo{person}{Alan Fekete}, \bibinfo{person}{Ali Ghodsi}, \bibinfo{person}{Joseph~M Hellerstein}, {and} \bibinfo{person}{Ion Stoica}.} \bibinfo{year}{2016}\natexlab{}.
\newblock \showarticletitle{Scalable atomic visibility with RAMP transactions}.
\newblock \bibinfo{journal}{\emph{ACM Transactions on Database Systems (TODS)}} \bibinfo{volume}{41}, \bibinfo{number}{3} (\bibinfo{year}{2016}), \bibinfo{pages}{1--45}.
\newblock
\urldef\tempurl%
\url{https://doi.org/10.1145/2909870}
\showDOI{\tempurl}


\bibitem[Berenson et~al\mbox{.}(1995)]%
        {si}
\bibfield{author}{\bibinfo{person}{Hal Berenson}, \bibinfo{person}{Philip~A. Bernstein}, \bibinfo{person}{Jim Gray}, \bibinfo{person}{Jim Melton}, \bibinfo{person}{Elizabeth~J. O'Neil}, {and} \bibinfo{person}{Patrick~E. O'Neil}.} \bibinfo{year}{1995}\natexlab{}.
\newblock \showarticletitle{A Critique of {ANSI} {SQL} Isolation Levels}. In \bibinfo{booktitle}{\emph{Proceedings of the 1995 {ACM} {SIGMOD} International Conference on Management of Data}}, \bibfield{editor}{\bibinfo{person}{Michael~J. Carey} {and} \bibinfo{person}{Donovan~A. Schneider}} (Eds.). \bibinfo{publisher}{{ACM} Press}, \bibinfo{pages}{1--10}.
\newblock
\urldef\tempurl%
\url{https://doi.org/10.1145/223784.223785}
\showDOI{\tempurl}


\bibitem[Bernstein and Goodman(1981)]%
        {bernstein1981concurrency}
\bibfield{author}{\bibinfo{person}{Philip~A Bernstein} {and} \bibinfo{person}{Nathan Goodman}.} \bibinfo{year}{1981}\natexlab{}.
\newblock \showarticletitle{Concurrency control in distributed database systems}.
\newblock \bibinfo{journal}{\emph{ACM Computing Surveys (CSUR)}} \bibinfo{volume}{13}, \bibinfo{number}{2} (\bibinfo{year}{1981}), \bibinfo{pages}{185--221}.
\newblock
\urldef\tempurl%
\url{https://doi.org/10.1145/356842.356846}
\showDOI{\tempurl}


\bibitem[Biswas and Enea(2019)]%
        {DBLP:journals/pacmpl/BiswasE19}
\bibfield{author}{\bibinfo{person}{Ranadeep Biswas} {and} \bibinfo{person}{Constantin Enea}.} \bibinfo{year}{2019}\natexlab{}.
\newblock \showarticletitle{On the complexity of checking transactional consistency}.
\newblock \bibinfo{journal}{\emph{Proc. {ACM} Program. Lang.}} \bibinfo{volume}{3}, \bibinfo{number}{{OOPSLA}} (\bibinfo{year}{2019}), \bibinfo{pages}{165:1--165:28}.
\newblock
\urldef\tempurl%
\url{https://doi.org/10.1145/3360591}
\showDOI{\tempurl}


\bibitem[Bravo et~al\mbox{.}(2021)]%
        {UniStore:ATC2021}
\bibfield{author}{\bibinfo{person}{Manuel Bravo}, \bibinfo{person}{Alexey Gotsman}, \bibinfo{person}{Borja de R{\'{e}}gil}, {and} \bibinfo{person}{Hengfeng Wei}.} \bibinfo{year}{2021}\natexlab{}.
\newblock \showarticletitle{UniStore: {A} fault-tolerant marriage of causal and strong consistency}. In \bibinfo{booktitle}{\emph{Proceedings of the 2021 {USENIX} Annual Technical Conference, {USENIX} {ATC} 2021}}, \bibfield{editor}{\bibinfo{person}{Irina Calciu} {and} \bibinfo{person}{Geoff Kuenning}} (Eds.). \bibinfo{publisher}{{USENIX} Association}, \bibinfo{pages}{923--937}.
\newblock
\urldef\tempurl%
\url{https://www.usenix.org/conference/atc21/presentation/bravo}
\showURL{%
\tempurl}


\bibitem[Burckhardt et~al\mbox{.}(2015)]%
        {pc}
\bibfield{author}{\bibinfo{person}{Sebastian Burckhardt}, \bibinfo{person}{Daan Leijen}, \bibinfo{person}{Jonathan Protzenko}, {and} \bibinfo{person}{Manuel F{\"{a}}hndrich}.} \bibinfo{year}{2015}\natexlab{}.
\newblock \showarticletitle{Global Sequence Protocol: {A} Robust Abstraction for Replicated Shared State}. In \bibinfo{booktitle}{\emph{29th European Conference on Object-Oriented Programming, {ECOOP} 2015}} \emph{(\bibinfo{series}{LIPIcs}, Vol.~\bibinfo{volume}{37})}. \bibinfo{publisher}{Schloss Dagstuhl - Leibniz-Zentrum f{\"{u}}r Informatik}, \bibinfo{pages}{568--590}.
\newblock
\urldef\tempurl%
\url{https://doi.org/10.4230/LIPICS.ECOOP.2015.568}
\showDOI{\tempurl}


\bibitem[Cerone et~al\mbox{.}(2015)]%
        {DBLP:conf/concur/Cerone0G15}
\bibfield{author}{\bibinfo{person}{Andrea Cerone}, \bibinfo{person}{Giovanni Bernardi}, {and} \bibinfo{person}{Alexey Gotsman}.} \bibinfo{year}{2015}\natexlab{}.
\newblock \showarticletitle{A Framework for Transactional Consistency Models with Atomic Visibility}. In \bibinfo{booktitle}{\emph{26th International Conference on Concurrency Theory, {CONCUR} 2015}} \emph{(\bibinfo{series}{LIPIcs}, Vol.~\bibinfo{volume}{42})}, \bibfield{editor}{\bibinfo{person}{Luca Aceto} {and} \bibinfo{person}{David de~Frutos{-}Escrig}} (Eds.). \bibinfo{publisher}{Schloss Dagstuhl - Leibniz-Zentrum f{\"{u}}r Informatik}, \bibinfo{pages}{58--71}.
\newblock
\urldef\tempurl%
\url{https://doi.org/10.4230/LIPICS.CONCUR.2015.58}
\showDOI{\tempurl}


\bibitem[Cheng et~al\mbox{.}(2021)]%
        {RAMP-TAO}
\bibfield{author}{\bibinfo{person}{Audrey Cheng}, \bibinfo{person}{Xiao Shi}, \bibinfo{person}{Lu Pan}, \bibinfo{person}{Anthony Simpson}, \bibinfo{person}{Neil Wheaton}, \bibinfo{person}{Shilpa Lawande}, \bibinfo{person}{Nathan Bronson}, \bibinfo{person}{Peter Bailis}, \bibinfo{person}{Natacha Crooks}, {and} \bibinfo{person}{Ion Stoica}.} \bibinfo{year}{2021}\natexlab{}.
\newblock \showarticletitle{{RAMP-TAO:} Layering Atomic Transactions on Facebook's Online {TAO} Data Store}.
\newblock \bibinfo{journal}{\emph{Proc. {VLDB} Endow.}} \bibinfo{volume}{14}, \bibinfo{number}{12} (\bibinfo{year}{2021}), \bibinfo{pages}{3014--3027}.
\newblock
\urldef\tempurl%
\url{https://doi.org/10.14778/3476311.3476379}
\showDOI{\tempurl}


\bibitem[Clavel et~al\mbox{.}(2007)]%
        {AllAboutMaude:Book2007}
\bibfield{author}{\bibinfo{person}{Manuel Clavel}, \bibinfo{person}{Francisco Dur\'{a}n}, \bibinfo{person}{Steven Eker}, \bibinfo{person}{Patrick Lincoln}, \bibinfo{person}{Narciso Mart\'{\i}-Oliet}, \bibinfo{person}{Jos\'{e} Meseguer}, {and} \bibinfo{person}{Carolyn Talcott}.} \bibinfo{year}{2007}\natexlab{}.
\newblock \bibinfo{booktitle}{\emph{All about Maude - a High-Performance Logical Framework: How to Specify, Program and Verify Systems in Rewriting Logic}}.
\newblock \bibinfo{publisher}{Springer-Verlag}, \bibinfo{address}{Berlin, Heidelberg}.
\newblock
\showISBNx{3540719407}


\bibitem[Corbett et~al\mbox{.}(2013)]%
        {corbett2013spanner}
\bibfield{author}{\bibinfo{person}{James~C. Corbett}, \bibinfo{person}{Jeffrey Dean}, \bibinfo{person}{Michael Epstein}, \bibinfo{person}{Andrew Fikes}, \bibinfo{person}{Christopher Frost}, \bibinfo{person}{J.~J. Furman}, \bibinfo{person}{Sanjay Ghemawat}, \bibinfo{person}{Andrey Gubarev}, \bibinfo{person}{Christopher Heiser}, \bibinfo{person}{Peter Hochschild}, \bibinfo{person}{Wilson~C. Hsieh}, \bibinfo{person}{Sebastian Kanthak}, \bibinfo{person}{Eugene Kogan}, \bibinfo{person}{Hongyi Li}, \bibinfo{person}{Alexander Lloyd}, \bibinfo{person}{Sergey Melnik}, \bibinfo{person}{David Mwaura}, \bibinfo{person}{David Nagle}, \bibinfo{person}{Sean Quinlan}, \bibinfo{person}{Rajesh Rao}, \bibinfo{person}{Lindsay Rolig}, \bibinfo{person}{Yasushi Saito}, \bibinfo{person}{Michal Szymaniak}, \bibinfo{person}{Christopher Taylor}, \bibinfo{person}{Ruth Wang}, {and} \bibinfo{person}{Dale Woodford}.} \bibinfo{year}{2013}\natexlab{}.
\newblock \showarticletitle{Spanner: Google's Globally Distributed Database}.
\newblock \bibinfo{journal}{\emph{{ACM} Transactions on Computer Systems (TOCS)}} \bibinfo{volume}{31}, \bibinfo{number}{3} (\bibinfo{year}{2013}), \bibinfo{pages}{8:1--8:22}.
\newblock
\urldef\tempurl%
\url{https://doi.org/10.1145/2491245}
\showDOI{\tempurl}


\bibitem[Crooks et~al\mbox{.}(2017)]%
        {seeing}
\bibfield{author}{\bibinfo{person}{Natacha Crooks}, \bibinfo{person}{Youer Pu}, \bibinfo{person}{Lorenzo Alvisi}, {and} \bibinfo{person}{Allen Clement}.} \bibinfo{year}{2017}\natexlab{}.
\newblock \showarticletitle{Seeing is Believing: A Client-Centric Specification of Database Isolation}. In \bibinfo{booktitle}{\emph{Proceedings of the ACM Symposium on Principles of Distributed Computing}} \emph{(\bibinfo{series}{PODC'17})}. \bibinfo{publisher}{ACM}, \bibinfo{pages}{73–82}.
\newblock


\bibitem[Davis et~al\mbox{.}(2020)]%
        {mongoDB-TLA+}
\bibfield{author}{\bibinfo{person}{A.~Jesse~Jiryu Davis}, \bibinfo{person}{Max Hirschhorn}, {and} \bibinfo{person}{Judah Schvimer}.} \bibinfo{year}{2020}\natexlab{}.
\newblock \showarticletitle{Extreme Modelling in Practice}.
\newblock \bibinfo{journal}{\emph{Proc. VLDB Endow.}} \bibinfo{volume}{13}, \bibinfo{number}{9} (\bibinfo{date}{May} \bibinfo{year}{2020}), \bibinfo{pages}{1346–1358}.
\newblock
\showISSN{2150-8097}
\urldef\tempurl%
\url{https://doi.org/10.14778/3397230.3397233}
\showDOI{\tempurl}


\bibitem[Didona et~al\mbox{.}(2018)]%
        {FriendFoe:VLDB2018}
\bibfield{author}{\bibinfo{person}{Diego Didona}, \bibinfo{person}{Rachid Guerraoui}, \bibinfo{person}{Jingjing Wang}, {and} \bibinfo{person}{Willy Zwaenepoel}.} \bibinfo{year}{2018}\natexlab{}.
\newblock \showarticletitle{Causal Consistency and Latency Optimality: Friend or Foe?}
\newblock \bibinfo{journal}{\emph{Proc. {VLDB} Endow.}} \bibinfo{volume}{11}, \bibinfo{number}{11} (\bibinfo{year}{2018}), \bibinfo{pages}{1618--1632}.
\newblock


\bibitem[Dou et~al\mbox{.}(2023)]%
        {Troc:ICSE2023}
\bibfield{author}{\bibinfo{person}{Wensheng Dou}, \bibinfo{person}{Ziyu Cui}, \bibinfo{person}{Qianwang Dai}, \bibinfo{person}{Jiansen Song}, \bibinfo{person}{Dong Wang}, \bibinfo{person}{Yu Gao}, \bibinfo{person}{Wei Wang}, \bibinfo{person}{Jun Wei}, \bibinfo{person}{Lei Chen}, \bibinfo{person}{Hanmo Wang}, \bibinfo{person}{Hua Zhong}, {and} \bibinfo{person}{Tao Huang}.} \bibinfo{year}{2023}\natexlab{}.
\newblock \showarticletitle{Detecting Isolation Bugs via Transaction Oracle Construction}. In \bibinfo{booktitle}{\emph{45th {IEEE/ACM} International Conference on Software Engineering, {ICSE} 2023}}. \bibinfo{publisher}{{IEEE}}, \bibinfo{pages}{1123--1135}.
\newblock
\urldef\tempurl%
\url{https://doi.org/10.1109/ICSE48619.2023.00101}
\showDOI{\tempurl}


\bibitem[{ElectricSQL}(2024)]%
        {ElectricSQL}
\bibfield{author}{\bibinfo{person}{{ElectricSQL}}.} \bibinfo{year}{2024}\natexlab{}.
\newblock
\newblock
\urldef\tempurl%
\url{https://electric-sql.com/}
\showURL{%
Retrieved March~7, 2025 from \tempurl}


\bibitem[Ghasemirad et~al\mbox{.}(2025)]%
        {GhasemiradSprengerLiu+-TACAS25}
\bibfield{author}{\bibinfo{person}{Shabnam Ghasemirad}, \bibinfo{person}{Christoph Sprenger}, \bibinfo{person}{Si Liu}, \bibinfo{person}{Luca Multazzu}, {and} \bibinfo{person}{David Basin}.} \bibinfo{year}{2025}\natexlab{}.
\newblock \showarticletitle{Pushing the Limit: Verified Performance-Optimal Causally-Consistent Database Transactions}. In \bibinfo{booktitle}{\emph{Tools and Algorithms for the Construction and Analysis of Systems - 31st International Conference, {TACAS} 2025, Held as Part of the European Joint Conferences on Theory and Practice of Software, {ETAPS} 2025}} \emph{(\bibinfo{series}{Lecture Notes in Computer Science})}. \bibinfo{publisher}{Springer}.
\newblock
\newblock
\shownote{{To} appear}.


\bibitem[Gu et~al\mbox{.}(2024)]%
        {isovista}
\bibfield{author}{\bibinfo{person}{Long Gu}, \bibinfo{person}{Si Liu}, \bibinfo{person}{Tiancheng Xing}, \bibinfo{person}{Hengfeng Wei}, \bibinfo{person}{Yuxing Chen}, {and} \bibinfo{person}{David Basin}.} \bibinfo{year}{2024}\natexlab{}.
\newblock \showarticletitle{IsoVista: Black-box Checking Database Isolation Guarantees}.
\newblock \bibinfo{journal}{\emph{Proc. {VLDB} Endow.}} \bibinfo{volume}{17}, \bibinfo{number}{12} (\bibinfo{year}{2024}), \bibinfo{pages}{4325--4328}.
\newblock
\urldef\tempurl%
\url{https://doi.org/10.14778/3685800.3685866}
\showDOI{\tempurl}


\bibitem[Hsieh(2017)]%
        {CockroachDB}
\bibfield{author}{\bibinfo{person}{Diana Hsieh}.} \bibinfo{year}{2017}\natexlab{}.
\newblock \bibinfo{title}{CockroachDB beta passes Jepsen testing}.
\newblock
\newblock
\urldef\tempurl%
\url{https://www.cockroachlabs.com/blog/cockroachdb-beta-passes-jepsen-testing/}
\showURL{%
Retrieved March~7, 2025 from \tempurl}


\bibitem[Huang et~al\mbox{.}(2023)]%
        {polysi}
\bibfield{author}{\bibinfo{person}{Kaile Huang}, \bibinfo{person}{Si Liu}, \bibinfo{person}{Zhenge Chen}, \bibinfo{person}{Hengfeng Wei}, \bibinfo{person}{David Basin}, \bibinfo{person}{Haixiang Li}, {and} \bibinfo{person}{Anqun Pan}.} \bibinfo{year}{2023}\natexlab{}.
\newblock \showarticletitle{Efficient Black-box Checking of Snapshot Isolation in Databases}.
\newblock \bibinfo{journal}{\emph{Proc. {VLDB} Endow.}} \bibinfo{volume}{16}, \bibinfo{number}{6} (\bibinfo{year}{2023}), \bibinfo{pages}{1264--1276}.
\newblock
\urldef\tempurl%
\url{https://doi.org/10.14778/3583140.3583145}
\showDOI{\tempurl}


\bibitem[Jepsen(2024)]%
        {jepsen-analyses}
\bibfield{author}{\bibinfo{person}{Jepsen}.} \bibinfo{year}{2024}\natexlab{}.
\newblock \bibinfo{title}{{Jepsen Analyses}}.
\newblock
\newblock
\urldef\tempurl%
\url{https://jepsen.io/analyses}
\showURL{%
Retrieved March~7, 2025 from \tempurl}


\bibitem[Jiang et~al\mbox{.}(2023)]%
        {txcheck}
\bibfield{author}{\bibinfo{person}{Zu{-}Ming Jiang}, \bibinfo{person}{Si Liu}, \bibinfo{person}{Manuel Rigger}, {and} \bibinfo{person}{Zhendong Su}.} \bibinfo{year}{2023}\natexlab{}.
\newblock \showarticletitle{Detecting Transactional Bugs in Database Engines via Graph-Based Oracle Construction}. In \bibinfo{booktitle}{\emph{17th {USENIX} Symposium on Operating Systems Design and Implementation, {OSDI} 2023}}, \bibfield{editor}{\bibinfo{person}{Roxana Geambasu} {and} \bibinfo{person}{Ed~Nightingale}} (Eds.). \bibinfo{publisher}{{USENIX} Association}, \bibinfo{pages}{397--417}.
\newblock
\urldef\tempurl%
\url{https://www.usenix.org/conference/osdi23/presentation/jiang}
\showURL{%
\tempurl}


\bibitem[Kingsbury(2018)]%
        {Dgraph}
\bibfield{author}{\bibinfo{person}{Kyle Kingsbury}.} \bibinfo{year}{2018}\natexlab{}.
\newblock \bibinfo{title}{Dgraph 1.0.2}.
\newblock
\newblock
\urldef\tempurl%
\url{https://jepsen.io/analyses/dgraph-1-0-2}
\showURL{%
Retrieved March~7, 2025 from \tempurl}


\bibitem[Kingsbury(2019)]%
        {YugaByte}
\bibfield{author}{\bibinfo{person}{Kyle Kingsbury}.} \bibinfo{year}{2019}\natexlab{}.
\newblock \bibinfo{title}{YugaByte DB 1.3.1}.
\newblock
\newblock
\urldef\tempurl%
\url{https://jepsen.io/analyses/yugabyte-db-1.3.1}
\showURL{%
Retrieved March~7, 2025 from \tempurl}


\bibitem[Kingsbury and Alvaro(2020)]%
        {elle}
\bibfield{author}{\bibinfo{person}{Kyle Kingsbury} {and} \bibinfo{person}{Peter Alvaro}.} \bibinfo{year}{2020}\natexlab{}.
\newblock \showarticletitle{Elle: Inferring Isolation Anomalies from Experimental Observations}.
\newblock \bibinfo{journal}{\emph{Proc. VLDB Endow.}} \bibinfo{volume}{14}, \bibinfo{number}{3} (\bibinfo{year}{2020}), \bibinfo{pages}{268–280}.
\newblock


\bibitem[Lesani et~al\mbox{.}(2016)]%
        {LesaniBC16}
\bibfield{author}{\bibinfo{person}{Mohsen Lesani}, \bibinfo{person}{Christian~J. Bell}, {and} \bibinfo{person}{Adam Chlipala}.} \bibinfo{year}{2016}\natexlab{}.
\newblock \showarticletitle{Chapar: certified causally consistent distributed key-value stores}. In \bibinfo{booktitle}{\emph{Proceedings of the 43rd Annual {ACM} {SIGPLAN-SIGACT} Symposium on Principles of Programming Languages, {POPL} 2016}}, \bibfield{editor}{\bibinfo{person}{Rastislav Bod{\'{\i}}k} {and} \bibinfo{person}{Rupak Majumdar}} (Eds.). \bibinfo{publisher}{{ACM}}, \bibinfo{pages}{357--370}.
\newblock
\urldef\tempurl%
\url{https://doi.org/10.1145/2837614.2837622}
\showDOI{\tempurl}


\bibitem[Library(2023)]%
        {galera-issue-1}
\bibfield{author}{\bibinfo{person}{Galera~Cluster Library}.} \bibinfo{year}{2023}\natexlab{}.
\newblock \bibinfo{title}{{Remove SNAPSHOT ISOLATION mention from FAQ page \#349}}.
\newblock
\newblock
\urldef\tempurl%
\url{https://github.com/codership/documentation/commit/cc8d6125f1767493eb61e2cc82f5a365ecee6e7a}
\showURL{%
Retrieved March~7, 2025 from \tempurl}


\bibitem[Lipton(1975)]%
        {DBLP:journals/cacm/Lipton75}
\bibfield{author}{\bibinfo{person}{Richard~J. Lipton}.} \bibinfo{year}{1975}\natexlab{}.
\newblock \showarticletitle{Reduction: {A} Method of Proving Properties of Parallel Programs}.
\newblock \bibinfo{journal}{\emph{Commun. {ACM}}} \bibinfo{volume}{18}, \bibinfo{number}{12} (\bibinfo{year}{1975}), \bibinfo{pages}{717--721}.
\newblock
\urldef\tempurl%
\url{https://doi.org/10.1145/361227.361234}
\showDOI{\tempurl}


\bibitem[Liu(2022)]%
        {lora}
\bibfield{author}{\bibinfo{person}{Si Liu}.} \bibinfo{year}{2022}\natexlab{}.
\newblock \showarticletitle{All in One: Design, Verification, and Implementation of SNOW-optimal Read Atomic Transactions}.
\newblock \bibinfo{journal}{\emph{{ACM} Trans. Softw. Eng. Methodol.}} \bibinfo{volume}{31}, \bibinfo{number}{3} (\bibinfo{year}{2022}), \bibinfo{pages}{43:1--43:44}.
\newblock
\urldef\tempurl%
\url{https://doi.org/10.1145/3494517}
\showDOI{\tempurl}


\bibitem[Liu et~al\mbox{.}(2024a)]%
        {plume}
\bibfield{author}{\bibinfo{person}{Si Liu}, \bibinfo{person}{Long Gu}, \bibinfo{person}{Hengfeng Wei}, {and} \bibinfo{person}{David Basin}.} \bibinfo{year}{2024}\natexlab{a}.
\newblock \showarticletitle{Plume: Efficient and Complete Black-Box Checking of Weak Isolation Levels}.
\newblock \bibinfo{journal}{\emph{Proc. {ACM} Program. Lang.}} \bibinfo{volume}{8}, \bibinfo{number}{{OOPSLA}} (\bibinfo{year}{2024}), \bibinfo{pages}{876--904}.
\newblock
\urldef\tempurl%
\url{https://doi.org/10.1145/3689742}
\showDOI{\tempurl}


\bibitem[Liu et~al\mbox{.}(2024b)]%
        {noc-noc}
\bibfield{author}{\bibinfo{person}{Si Liu}, \bibinfo{person}{Luca Multazzu}, \bibinfo{person}{Hengfeng Wei}, {and} \bibinfo{person}{David Basin}.} \bibinfo{year}{2024}\natexlab{b}.
\newblock \showarticletitle{{NOC-NOC:} Towards Performance-optimal Distributed Transactions}.
\newblock \bibinfo{journal}{\emph{Proc. {ACM} Manag. Data}} \bibinfo{volume}{2}, \bibinfo{number}{1} (\bibinfo{year}{2024}), \bibinfo{pages}{9:1--9:25}.
\newblock
\urldef\tempurl%
\url{https://doi.org/10.1145/3639264}
\showDOI{\tempurl}


\bibitem[Liu et~al\mbox{.}(2019a)]%
        {ua}
\bibfield{author}{\bibinfo{person}{Si Liu}, \bibinfo{person}{Peter~Csaba {\"{O}}lveczky}, \bibinfo{person}{Qi Wang}, \bibinfo{person}{Indranil Gupta}, {and} \bibinfo{person}{Jos{\'{e}} Meseguer}.} \bibinfo{year}{2019}\natexlab{a}.
\newblock \showarticletitle{Read atomic transactions with prevention of lost updates: {ROLA} and its formal analysis}.
\newblock \bibinfo{journal}{\emph{Formal Aspects Comput.}} \bibinfo{volume}{31}, \bibinfo{number}{5} (\bibinfo{year}{2019}), \bibinfo{pages}{503--540}.
\newblock
\urldef\tempurl%
\url{https://doi.org/10.1007/S00165-019-00489-W}
\showDOI{\tempurl}


\bibitem[Liu et~al\mbox{.}(2019b)]%
        {DBLP:conf/tacas/LiuOZWM19}
\bibfield{author}{\bibinfo{person}{Si Liu}, \bibinfo{person}{Peter~Csaba {\"{O}}lveczky}, \bibinfo{person}{Min Zhang}, \bibinfo{person}{Qi Wang}, {and} \bibinfo{person}{Jos{\'{e}} Meseguer}.} \bibinfo{year}{2019}\natexlab{b}.
\newblock \showarticletitle{Automatic Analysis of Consistency Properties of Distributed Transaction Systems in Maude}. In \bibinfo{booktitle}{\emph{Tools and Algorithms for the Construction and Analysis of Systems - 25th International Conference, {TACAS} 2019, Held as Part of the European Joint Conferences on Theory and Practice of Software, {ETAPS} 2019}} \emph{(\bibinfo{series}{Lecture Notes in Computer Science}, Vol.~\bibinfo{volume}{11428})}, \bibfield{editor}{\bibinfo{person}{Tom{\'{a}}s Vojnar} {and} \bibinfo{person}{Lijun Zhang}} (Eds.). \bibinfo{publisher}{Springer}, \bibinfo{pages}{40--57}.
\newblock
\urldef\tempurl%
\url{https://doi.org/10.1007/978-3-030-17465-1\_3}
\showDOI{\tempurl}


\bibitem[Lloyd et~al\mbox{.}(2013)]%
        {Eiger:NSDI2013}
\bibfield{author}{\bibinfo{person}{Wyatt Lloyd}, \bibinfo{person}{Michael~J. Freedman}, \bibinfo{person}{Michael Kaminsky}, {and} \bibinfo{person}{David~G. Andersen}.} \bibinfo{year}{2013}\natexlab{}.
\newblock \showarticletitle{Stronger Semantics for Low-Latency Geo-Replicated Storage}. In \bibinfo{booktitle}{\emph{Proceedings of the 10th {USENIX} Symposium on Networked Systems Design and Implementation, {NSDI} 2013}}, \bibfield{editor}{\bibinfo{person}{Nick Feamster} {and} \bibinfo{person}{Jeffrey~C. Mogul}} (Eds.). \bibinfo{publisher}{{USENIX} Association}, \bibinfo{pages}{313--328}.
\newblock
\urldef\tempurl%
\url{https://www.usenix.org/conference/nsdi13/technical-sessions/presentation/lloyd}
\showURL{%
\tempurl}


\bibitem[Lu et~al\mbox{.}(2016)]%
        {SNOW:OSDI2016}
\bibfield{author}{\bibinfo{person}{Haonan Lu}, \bibinfo{person}{Christopher Hodsdon}, \bibinfo{person}{Khiem Ngo}, \bibinfo{person}{Shuai Mu}, {and} \bibinfo{person}{Wyatt Lloyd}.} \bibinfo{year}{2016}\natexlab{}.
\newblock \showarticletitle{The {SNOW} Theorem and Latency-Optimal Read-Only Transactions}. In \bibinfo{booktitle}{\emph{12th {USENIX} Symposium on Operating Systems Design and Implementation, {OSDI} 2016}}, \bibfield{editor}{\bibinfo{person}{Kimberly Keeton} {and} \bibinfo{person}{Timothy Roscoe}} (Eds.). \bibinfo{publisher}{{USENIX} Association}, \bibinfo{pages}{135--150}.
\newblock
\urldef\tempurl%
\url{https://www.usenix.org/conference/osdi16/technical-sessions/presentation/lu}
\showURL{%
\tempurl}


\bibitem[Lu et~al\mbox{.}(2023)]%
        {osdi23}
\bibfield{author}{\bibinfo{person}{Haonan Lu}, \bibinfo{person}{Shuai Mu}, \bibinfo{person}{Siddhartha Sen}, {and} \bibinfo{person}{Wyatt Lloyd}.} \bibinfo{year}{2023}\natexlab{}.
\newblock \showarticletitle{{NCC}: Natural Concurrency Control for Strictly Serializable Datastores by Avoiding the {Timestamp-Inversion} Pitfall}. In \bibinfo{booktitle}{\emph{17th USENIX Symposium on Operating Systems Design and Implementation (OSDI 23)}}. \bibinfo{publisher}{USENIX Association}, \bibinfo{pages}{305--323}.
\newblock


\bibitem[Lynch and Vaandrager(1995)]%
        {DBLP:journals/iandc/LynchV95}
\bibfield{author}{\bibinfo{person}{Nancy~A. Lynch} {and} \bibinfo{person}{Frits~W. Vaandrager}.} \bibinfo{year}{1995}\natexlab{}.
\newblock \showarticletitle{Forward and Backward Simulations: I. Untimed Systems}.
\newblock \bibinfo{journal}{\emph{Inf. Comput.}} \bibinfo{volume}{121}, \bibinfo{number}{2} (\bibinfo{year}{1995}), \bibinfo{pages}{214--233}.
\newblock
\urldef\tempurl%
\url{https://doi.org/10.1006/inco.1995.1134}
\showDOI{\tempurl}


\bibitem[Mehdi et~al\mbox{.}(2017)]%
        {Slowdown:NSDI2017}
\bibfield{author}{\bibinfo{person}{Syed~Akbar Mehdi}, \bibinfo{person}{Cody Littley}, \bibinfo{person}{Natacha Crooks}, \bibinfo{person}{Lorenzo Alvisi}, \bibinfo{person}{Nathan Bronson}, {and} \bibinfo{person}{Wyatt Lloyd}.} \bibinfo{year}{2017}\natexlab{}.
\newblock \showarticletitle{I Can't Believe It's Not Causal! Scalable Causal Consistency with No Slowdown Cascades}. In \bibinfo{booktitle}{\emph{14th {USENIX} Symposium on Networked Systems Design and Implementation, {NSDI} 2017}}, \bibfield{editor}{\bibinfo{person}{Aditya Akella} {and} \bibinfo{person}{Jon Howell}} (Eds.). \bibinfo{publisher}{{USENIX} Association}, \bibinfo{pages}{453--468}.
\newblock
\urldef\tempurl%
\url{https://www.usenix.org/conference/nsdi17/technical-sessions/presentation/mehdi}
\showURL{%
\tempurl}


\bibitem[Microsoft(2024)]%
        {CosmosDB}
\bibfield{author}{\bibinfo{person}{Microsoft}.} \bibinfo{year}{2024}\natexlab{}.
\newblock \bibinfo{title}{{Consistency levels in Azure Cosmos DB}}.
\newblock
\newblock
\urldef\tempurl%
\url{https://learn.microsoft.com/en-us/azure/cosmos-db/consistency-levels}
\showURL{%
Retrieved March~7, 2025 from \tempurl}


\bibitem[{MongoDB}(2024)]%
        {MongoDB}
\bibfield{author}{\bibinfo{person}{{MongoDB}}.} \bibinfo{year}{2024}\natexlab{}.
\newblock
\newblock
\urldef\tempurl%
\url{https://www.mongodb.com/}
\showURL{%
Retrieved March~7, 2025 from \tempurl}


\bibitem[Neo4j(2024)]%
        {Neo4j}
\bibfield{author}{\bibinfo{person}{Neo4j}.} \bibinfo{year}{2024}\natexlab{}.
\newblock
\newblock
\urldef\tempurl%
\url{https://neo4j.com/}
\showURL{%
Retrieved March~7, 2025 from \tempurl}


\bibitem[Newcombe et~al\mbox{.}(2015)]%
        {amazon}
\bibfield{author}{\bibinfo{person}{Chris Newcombe}, \bibinfo{person}{Tim Rath}, \bibinfo{person}{Fan Zhang}, \bibinfo{person}{Bogdan Munteanu}, \bibinfo{person}{Marc Brooker}, {and} \bibinfo{person}{Michael Deardeuff}.} \bibinfo{year}{2015}\natexlab{}.
\newblock \showarticletitle{How Amazon web services uses formal methods}.
\newblock \bibinfo{journal}{\emph{Commun. {ACM}}} \bibinfo{volume}{58}, \bibinfo{number}{4} (\bibinfo{year}{2015}), \bibinfo{pages}{66--73}.
\newblock
\urldef\tempurl%
\url{https://doi.org/10.1145/2699417}
\showDOI{\tempurl}


\bibitem[Nipkow and Klein(2014)]%
        {nipkow2014concrete}
\bibfield{author}{\bibinfo{person}{Tobias Nipkow} {and} \bibinfo{person}{Gerwin Klein}.} \bibinfo{year}{2014}\natexlab{}.
\newblock \bibinfo{booktitle}{\emph{Concrete semantics: with Isabelle/HOL}}.
\newblock \bibinfo{publisher}{Springer}.
\newblock
\urldef\tempurl%
\url{https://doi.org/10.1007/978-3-319-10542-0}
\showDOI{\tempurl}


\bibitem[Nipkow et~al\mbox{.}(2002)]%
        {DBLP:books/sp/NipkowPW02}
\bibfield{author}{\bibinfo{person}{Tobias Nipkow}, \bibinfo{person}{Lawrence~C. Paulson}, {and} \bibinfo{person}{Markus Wenzel}.} \bibinfo{year}{2002}\natexlab{}.
\newblock \bibinfo{booktitle}{\emph{Isabelle/HOL - {A} Proof Assistant for Higher-Order Logic}}. \bibinfo{series}{Lecture Notes in Computer Science}, Vol.~\bibinfo{volume}{2283}.
\newblock \bibinfo{publisher}{Springer}.
\newblock
\showISBNx{3-540-43376-7}
\urldef\tempurl%
\url{https://doi.org/10.1007/3-540-45949-9}
\showDOI{\tempurl}


\bibitem[Papadimitriou(1979)]%
        {serializability}
\bibfield{author}{\bibinfo{person}{Christos~H Papadimitriou}.} \bibinfo{year}{1979}\natexlab{}.
\newblock \showarticletitle{The serializability of concurrent database updates}.
\newblock \bibinfo{journal}{\emph{Journal of the ACM (JACM)}} \bibinfo{volume}{26}, \bibinfo{number}{4} (\bibinfo{year}{1979}), \bibinfo{pages}{631--653}.
\newblock


\bibitem[Sovran et~al\mbox{.}(2011)]%
        {psi}
\bibfield{author}{\bibinfo{person}{Yair Sovran}, \bibinfo{person}{Russell Power}, \bibinfo{person}{Marcos~K. Aguilera}, {and} \bibinfo{person}{Jinyang Li}.} \bibinfo{year}{2011}\natexlab{}.
\newblock \showarticletitle{Transactional storage for geo-replicated systems}. In \bibinfo{booktitle}{\emph{Proceedings of the 23rd {ACM} Symposium on Operating Systems Principles 2011, {SOSP} 2011}}, \bibfield{editor}{\bibinfo{person}{Ted Wobber} {and} \bibinfo{person}{Peter Druschel}} (Eds.). \bibinfo{publisher}{{ACM}}, \bibinfo{pages}{385--400}.
\newblock
\urldef\tempurl%
\url{https://doi.org/10.1145/2043556.2043592}
\showDOI{\tempurl}


\bibitem[Spirovska et~al\mbox{.}(2021)]%
        {OCC:TPDS2021}
\bibfield{author}{\bibinfo{person}{Kristina Spirovska}, \bibinfo{person}{Diego Didona}, {and} \bibinfo{person}{Willy Zwaenepoel}.} \bibinfo{year}{2021}\natexlab{}.
\newblock \showarticletitle{Optimistic Causal Consistency for Geo-Replicated Key-Value Stores}.
\newblock \bibinfo{journal}{\emph{{IEEE} Trans. Parallel Distributed Syst.}} \bibinfo{volume}{32}, \bibinfo{number}{3} (\bibinfo{year}{2021}), \bibinfo{pages}{527--542}.
\newblock


\bibitem[Sprenger et~al\mbox{.}(2020)]%
        {SprengerKEWMCB2020}
\bibfield{author}{\bibinfo{person}{Christoph Sprenger}, \bibinfo{person}{Tobias Klenze}, \bibinfo{person}{Marco Eilers}, \bibinfo{person}{Felix~A. Wolf}, \bibinfo{person}{Peter M{\"{u}}ller}, \bibinfo{person}{Martin Clochard}, {and} \bibinfo{person}{David Basin}.} \bibinfo{year}{2020}\natexlab{}.
\newblock \showarticletitle{Igloo: soundly linking compositional refinement and separation logic for distributed system verification}.
\newblock \bibinfo{journal}{\emph{Proc. {ACM} Program. Lang.}} \bibinfo{volume}{4}, \bibinfo{number}{{OOPSLA}} (\bibinfo{year}{2020}), \bibinfo{pages}{152:1--152:31}.
\newblock
\urldef\tempurl%
\url{https://doi.org/10.1145/3428220}
\showDOI{\tempurl}


\bibitem[Systems(2024)]%
        {Apalache}
\bibfield{author}{\bibinfo{person}{Informal Systems}.} \bibinfo{year}{2024}\natexlab{}.
\newblock \bibinfo{title}{Apalache}.
\newblock
\newblock
\urldef\tempurl%
\url{https://apalache.informal.systems/}
\showURL{%
Retrieved March~7, 2025 from \tempurl}


\bibitem[Tan et~al\mbox{.}(2020)]%
        {cobra}
\bibfield{author}{\bibinfo{person}{Cheng Tan}, \bibinfo{person}{Changgeng Zhao}, \bibinfo{person}{Shuai Mu}, {and} \bibinfo{person}{Michael Walfish}.} \bibinfo{year}{2020}\natexlab{}.
\newblock \showarticletitle{Cobra: Making Transactional Key-Value Stores Verifiably Serializable}. In \bibinfo{booktitle}{\emph{14th {USENIX} Symposium on Operating Systems Design and Implementation, {OSDI} 2020}}. \bibinfo{publisher}{{USENIX} Association}, \bibinfo{pages}{63--80}.
\newblock
\urldef\tempurl%
\url{https://www.usenix.org/conference/osdi20/presentation/tan}
\showURL{%
\tempurl}


\bibitem[Wei et~al\mbox{.}(2015)]%
        {DrTM}
\bibfield{author}{\bibinfo{person}{Xingda Wei}, \bibinfo{person}{Jiaxin Shi}, \bibinfo{person}{Yanzhe Chen}, \bibinfo{person}{Rong Chen}, {and} \bibinfo{person}{Haibo Chen}.} \bibinfo{year}{2015}\natexlab{}.
\newblock \showarticletitle{Fast In-Memory Transaction Processing Using RDMA and HTM}. In \bibinfo{booktitle}{\emph{Proceedings of the 25th Symposium on Operating Systems Principles}} \emph{(\bibinfo{series}{SOSP '15})}. \bibinfo{publisher}{{ACM}}, \bibinfo{pages}{87–104}.
\newblock
\urldef\tempurl%
\url{https://doi.org/10.1145/2815400.2815419}
\showDOI{\tempurl}


\bibitem[Xiong et~al\mbox{.}(2020)]%
        {DBLP:conf/ecoop/XiongCRG19}
\bibfield{author}{\bibinfo{person}{Shale Xiong}, \bibinfo{person}{Andrea Cerone}, \bibinfo{person}{Azalea Raad}, {and} \bibinfo{person}{Philippa Gardner}.} \bibinfo{year}{2020}\natexlab{}.
\newblock \showarticletitle{Data Consistency in Transactional Storage Systems: {A} Centralised Semantics}. In \bibinfo{booktitle}{\emph{34th European Conference on Object-Oriented Programming, {ECOOP} 2020}} \emph{(\bibinfo{series}{LIPIcs}, Vol.~\bibinfo{volume}{166})}. \bibinfo{publisher}{Schloss Dagstuhl - Leibniz-Zentrum f{\"{u}}r Informatik}, \bibinfo{pages}{21:1--21:31}.
\newblock
\urldef\tempurl%
\url{https://doi.org/10.4230/LIPICS.ECOOP.2020.21}
\showDOI{\tempurl}


\bibitem[Zhang et~al\mbox{.}(2015)]%
        {tapir}
\bibfield{author}{\bibinfo{person}{Irene Zhang}, \bibinfo{person}{Naveen~Kr. Sharma}, \bibinfo{person}{Adriana Szekeres}, \bibinfo{person}{Arvind Krishnamurthy}, {and} \bibinfo{person}{Dan R.~K. Ports}.} \bibinfo{year}{2015}\natexlab{}.
\newblock \showarticletitle{Building consistent transactions with inconsistent replication}. In \bibinfo{booktitle}{\emph{Proceedings of the 25th Symposium on Operating Systems Principles, {SOSP} 2015}}, \bibfield{editor}{\bibinfo{person}{Ethan~L. Miller} {and} \bibinfo{person}{Steven Hand}} (Eds.). \bibinfo{publisher}{{ACM}}, \bibinfo{pages}{263--278}.
\newblock
\urldef\tempurl%
\url{https://doi.org/10.1145/2815400.2815404}
\showDOI{\tempurl}


\bibitem[Zhang et~al\mbox{.}(2018)]%
        {DBLP:journals/tocs/ZhangSSKP18}
\bibfield{author}{\bibinfo{person}{Irene Zhang}, \bibinfo{person}{Naveen~Kr. Sharma}, \bibinfo{person}{Adriana Szekeres}, \bibinfo{person}{Arvind Krishnamurthy}, {and} \bibinfo{person}{Dan R.~K. Ports}.} \bibinfo{year}{2018}\natexlab{}.
\newblock \showarticletitle{Building Consistent Transactions with Inconsistent Replication}.
\newblock \bibinfo{journal}{\emph{{ACM} Trans. Comput. Syst.}} \bibinfo{volume}{35}, \bibinfo{number}{4} (\bibinfo{year}{2018}), \bibinfo{pages}{12:1--12:37}.
\newblock
\urldef\tempurl%
\url{https://doi.org/10.1145/3269981}
\showDOI{\tempurl}


\bibitem[Zhang et~al\mbox{.}(2019)]%
        {tapir-code}
\bibfield{author}{\bibinfo{person}{Irene Zhang}, \bibinfo{person}{Naveen~Kr. Sharma}, \bibinfo{person}{Michael Whittaker}, {and} \bibinfo{person}{Maxime Caron}.} \bibinfo{year}{2019}\natexlab{}.
\newblock \bibinfo{title}{{TAPIR}}.
\newblock
\newblock
\urldef\tempurl%
\url{https://github.com/UWSysLab/tapir}
\showURL{%
Retrieved March~7, 2025 from \tempurl}


\end{thebibliography}

% uncomment to cut off appendix
%\end{document}

\iffull
%\newpage
\appendix
\section{Validating RA Violations in TAPIR}
\label{app:validation}

%overview
Using the black-box database isolation checker IsoVista~\cite{isovista},
we confirmed the violation of  \emph{atomic visibility} in TAPIR's implementation~\cite{tapir-code} that
complies with the specification in its journal version~\cite{DBLP:journals/tocs/ZhangSSKP18}. 
By treating TAPIR as a black box, we preserved its codebase, thereby minimizing the risk of introducing errors on our end.

\inlsec{Validation Process}
Our black-box validation of \emph{read atomicity} (RA) violations in TAPIR is conducted using the following  steps:

\begin{enumerate}
    \item TAPIR clients issue transactional requests generated by its workload generator;
    \item Each client session logs the requests it sends and the corresponding results returned by the database;
    \item  The logs from all the client sessions are merged into a unified history;
    \item IsoVista then analyzes this history to determine whether it satisfies RA.
\end{enumerate}

\noindent 
Specifically, in Step (4), IsoVista constructs a Biswas and Enea's dependency graph~\cite{DBLP:journals/pacmpl/BiswasE19} 
from the history, and
searches for specific subgraphs that represent various RA anomalies like \emph{fractured reads} (see also Section~\ref{subsec:iso}).

\inlsec{Experimental Setup}
We co-located the client threads (or
sessions) in a local machine 
 with an Intel Core i3-1005G1 CPU and 8GB memory.
We set the parameters of TAPIR's workload generator as:\footnote{Our fork of the TAPIR codebase, along with 
the experimental setup and result, is available at
\url{https://doi.org/10.5281/zenodo.14991194} (in the VerIso/TapirBBTest subfolder).}

\begin{lstlisting}[numbers=none]
nshard=2         # number of shards
nclient=3        # number of clients to run (per machine)
nkeys=2          # number of keys to use
rtime=1          # duration to run
tlen=2           # transaction length
wper=50          # writes percentage
err=0            # error
skew=0           # skew
zalpha=-1        # zipf alpha (-1 to enable uniform)
\end{lstlisting}

\begin{figure}[t]
\begin{center}
   \includegraphics[width=.8\columnwidth]{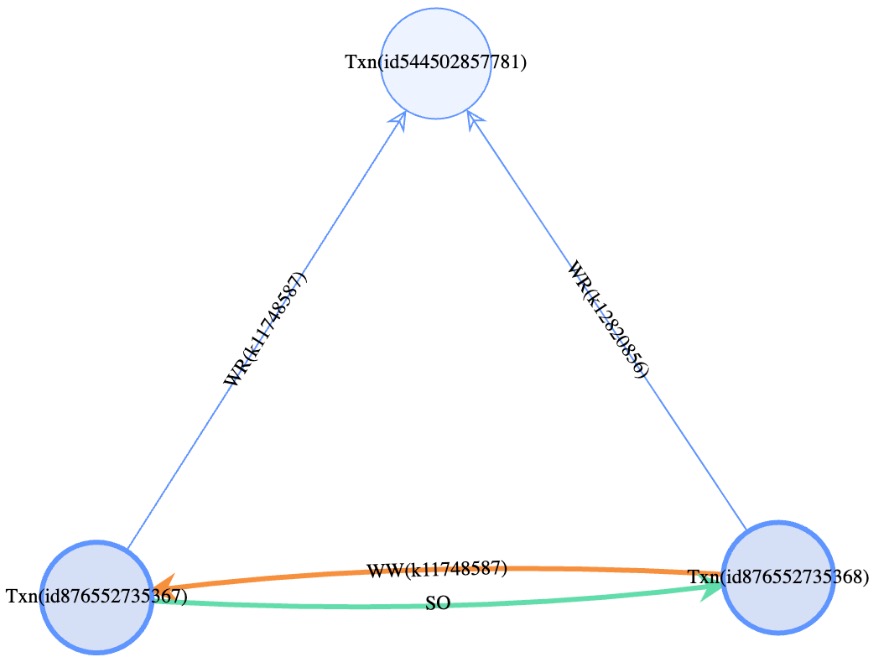}
\end{center}
   \caption{A counterexample of RA returned by IsoVista, which corresponds to the fracture reads anomaly.}
   \label{app:fig:bug}
\end{figure}

\inlsec{Counterexample}
Figure~\ref{app:fig:bug} shows a counterexample of RA, specifically a fractured reads anomaly, that has been detected, minimized, and  
visualized by IsoVista.
This counterexample involves three transactions from two clients: %Txn\_876552735367 
Txn(...67)
%(T1) 
and 
%Txn\_876552735368 
Txn(...68)
%(T2) 
from 
%Clt\_876552735270 
Clt(...70), 
%(C1), 
along with 
%Txn\_544502857781 
Txn(...81)
%(T3) 
from 
%Clt\_544502857701 
Clt(...01).\footnote{The client IDs can be seen in IsoVista by expanding the nodes in Figure~\ref{app:fig:bug}.} 
%(C2). 
It arises from conflicting dependencies between Txn(...67) and Txn(...68), represented by the cycle shown in Figure~\ref{app:fig:bug}. 
This cycle is built as follows:

\begin{itemize}
    \item Based on Biswas and Enea's theory~\cite{DBLP:journals/pacmpl/BiswasE19}, 
since (i) Txn(...81) reads Txn(...67)'s write to  %key 11748587,
Key(...87),
(ii) Txn(...81) reads Txn(...68)'s write to  
%key 12820856,
Key(...56)
and (iii) Txn(...68) also writes to %the key 11748587, 
Key(...87), 
IsoVista \emph{infers} a version order (or WW dependency) on  
%key 11748587
Key(...87)
from Txn(...68) to Txn(...67).

\item Since Txn(...67) is issued before Txn(...68) in the same session, there is a session order (or SO dependency) from Txn(...67) to Txn(...68). 
\end{itemize}

Intuitively, this counterexample exhibits fractured reads because 
Txn(...81) observes only part of Txn(...68)'s updates. Specifically, Txn(...68)'s write to %the key 12820856 
Key(...56)
is visible to Txn(...81), while its write to  %key 11748587
Key(...87)
is not; instead, Txn(...81) reads an older version of 
%the key 11748587 
Key(...87)
written by Txn(...67).

%T3
%r(11748587,619076138487,544502857701,544502857781)
%r(12820856,86471422672, 544502857701,544502857781)

%T1
%w(11748587,619076138487,876552735270,876552735367)
%w(12820856,751344751289,876552735270,876552735367)

%T2
%w(12820856,86471422672, 876552735270,876552735368)
%w(11748587,111363426644,876552735270,876552735368)
\fi

\end{document}